\documentclass[12pt,reqno]{amsart}
\usepackage{amsmath,amsfonts,amsthm,graphicx}
\usepackage{a4}

\newcommand{\dcrit}[1][]{D_\mathrm{crit}^\mathrm{#1}}
\newcommand{\R}{\mathbb{R}}
\newcommand{\rmd}{\mathrm{d}}
\newcommand{\ttg}[1]{\ensuremath{(\ref{#1})}}

\newtheorem{theorem}{Theorem}
\newtheorem{lemma}[theorem]{Lemma}

\begin{document}


\title{Unilateral regulation breaks regularity of Turing patterns%
}

\author{Tom\'a\v s Vejchodsk\'y, Filip Jaro\v{s}, Milan Ku\v{c}era, Vojt\v{e}ch Ryb\'a\v{r}}

\address{
  Tom{\'a}{\v s} Vejchodsk{\'y},
  Institute of Mathematics,
  Academy of Sciences,
  {\v Z}itn{\'a} 25, CZ-115 67 Praha 1,
  Czech Republic,
}\email{vejchod@math.cas.cz}
\address{
  Filip Jaro\v{s},
Faculty of Arts,
University of Hradec Kr\'alov\'e,
n\'am\v{e}st\'{\i} Svobody 331, 
CZ-500\,02 Hradec Kr\'alov\'e,  
Czech Republic.
}\email{filip.jaros@uhk.cz} 
\address{
  Milan Ku\v{c}era,
  Institute of Mathematics,
  Academy of Sciences,
  {\v Z}itn{\'a} 25, CZ-115 67 Praha 1,
  Czech Republic
and 
  Department of Mathematics, Faculty of Applied Sciences, 
  University of West Bohemia in Pilsen,
  Univerzitn\' i 8, 30614 Plze\v n, 
  Czech Republic.
}\email{kucera@math.cas.cz}
\address{
  Vojt\v{e}ch Ryb\'a\v{r},
  Institute of Mathematics,
  Academy of Sciences,
  {\v Z}itn{\'a} 25, CZ-115 67 Praha 1,
  Czech Republic.
}\email{rybar@math.cas.cz}

\thanks{Date: 22 August 2017.\\
\mbox{}\quad\;Published in Physical Review E {\bf 96}, 022212 (2017), DOI: 10.1103/PhysRevE.96.022212}%
%


\begin{abstract}
We consider a reaction-diffusion system undergoing Turing instability and augment it by an additional unilateral source term. We investigate its influence on the Turing instability and on the character of resulting patterns. The nonsmooth positively homogeneous unilateral term $\tau v^-$ has favourable properties, but the standard linear stability analysis cannot be performed. We illustrate the importance of the nonsmoothness by a numerical case study, which shows that the Turing instability can considerably change if we replace this term by its arbitrarily precise smooth approximation. However, the nonsmooth unilateral term and all its approximations yield qualitatively similar patterns although not necessarily developing from small disturbances of the spatially homogeneous steady state. Further, we show that the unilateral source breaks the approximate symmetry and regularity of the classical patterns and yields asymmetric and irregular patterns. Moreover, a given system with a unilateral source produces spatial patterns even for diffusion parameters with ratios closer to 1 than the same system without any unilateral term. 
\end{abstract}


%
\maketitle


\section{Introduction}
\label{se:intro}

Reaction-diffusion systems are frequently used to model the initiation of animal forms and patterns. After publication of Turing's purely theoretical paper \cite{Turing1952}, growing number of biologists succeeded in matching empirical data with mathematical simulations. Morphogens with Turing-like behaviour were found in the process of hair follicles formation
\cite{Mou2006}, 
the generation of transverse ridges of the palate
\cite{Economou2012} 
or patterning the germ layers
\cite{Chen2001,Schier2009}. 
The concept of reactions and diffusion of morphogens was widened to the interactions of pigment cells. In the case of zebrafish, the validity of this model was tested on individuals with ablated skin
\cite{Kondo2002,Kondo2010}. 
Turing's mechanism is also used to model the formation of coat patterns 
in mammals, see for example \cite{LiuLiaMai2006,Murray2003}.



We will introduce a unilateral term $\hat g(v)$ to the
concrete reaction-diffusion system introduced in \cite{BarVarAra1999} to model skin patterns in fish and used in \cite{LiuLiaMai2006} to model coat patterns of jaguar and leopard, see \ttg{eq:Maini+g}. This particular system will be provided in Section~\ref{se:smooth} and it fits into a general scheme
\begin{align}
  \label{eq:zdroj}
  \frac{\partial u }{\partial t} &= d_1 \Delta u + f(u,v) \quad\text{in }\Omega,\\ \nonumber
  \frac{\partial v }{\partial t} &= d_2 \Delta v + g(u,v) + \hat g(v)
  \quad\text{in }\Omega
\end{align}
with the usual homogeneous Neumann boundary conditions 
\begin{equation}
  \label{eq:bc1}
  \frac{\partial u}{\partial n} = \frac{\partial v}{\partial n} = 0
  \quad\text{on }\partial\Omega.
\end{equation}
This system models the diffusion and nonlinear interactions of two morphogens. 
The domain 
$\Omega \subset \R^2$ represents the tissue,
$t$ denotes the time variable,
$d_1$ and $d_2$ are diffusion coefficients,
smooth functions $f(u,v)$ and $g(u,v)$ describe interactions between the morphogens, 
and $n$ stands for the unit outward facing normal vector to the boundary $\partial\Omega$. 
In accordance with the original publications \cite{BarVarAra1999,LiuLiaMai2006}, 
the quantities $u$, $v$ denote deviations of morphogens concentrations $U$, $V$ from a positive spatially homogeneous equilibrium concentrations $\bar U$, $\bar V$. Thus, negative deviations $u = U - \bar U$ and $v = V - \bar V$ can still correspond to positive concentrations $U$ and $V$.
Values of $\bar U$ and $\bar V$ are not specified even in the original publications \cite{BarVarAra1999,LiuLiaMai2006}. In fact, they can be chosen arbitrarily and substitution $u = U - \bar U$ and $v = V - \bar V$ into \eqref{eq:zdroj} yields a system for concentrations $U$ and $V$. In all numerical calculations presented in this paper the deviations $u$ and $v$ lie within the range $(-1.5,1.5)$. Hence, if $\bar U >1.5$, $\bar V >1.5$ then concentrations $U=u+\bar U$, $V=v+\bar V$ remain positive.

%

A novelty in system \eqref{eq:zdroj} is the additional term $\hat g(v)$ which is unilateral in the sense that
there is a threshold value $\theta$ such that 
$$
\hat g(v) > 0 \mbox{ for } v < \theta \mbox{ and } \hat g(v) = 0 \mbox{ otherwise.}
$$
This term describes an additional source
active only if the concentration of the second morphogen decreases below the threshold $\theta$.
The key point is that the function $\hat g(v)$ can be nonsmooth, a typical example being 
$\hat g(v) = \tau (v-\theta)^-$, where 
$(v-\theta)^- = (|v-\theta| - v + \theta)/2$ stands for the negative part of $v - \theta$
and $\tau > 0$ controls the strength of this unilateral source.
An alternative example is the saturation term $\hat g(v) = \tau (1-\exp[-(v-\theta)^-])$ that models the limited ability of cells to produce morphogens.

We will have $f(0,0)=g(0,0)=0$. Thus, if $\hat g(0)= 0$ then $(u,v) = (0,0)$ will be a constant stationary solution of \ttg{eq:zdroj} with \ttg{eq:bc1}.
We will also consider cases when $\hat g(0)\ne 0$ and there exist
nontrivial $(\bar u, \bar v)$ satisfying 
$f(\bar u,\bar v)=g(\bar u,\bar v) +\hat g(\bar v) =0$.
Then $u=\bar u,\ v=\bar v$ is a constant stationary solution of
\ttg{eq:zdroj} with \ttg{eq:bc1}.
In both cases, we refer to this constant steady state $(\bar u, \bar v)$ as a \emph{ground state}.
We say that system \ttg{eq:zdroj} with \ttg{eq:bc1} undergoes the 
\emph{Turing diffusion driven instability} if
the ground state is stable 
with respect to small spatially \emph{homogeneous} perturbations and unstable 
with respect to small spatially \emph{nonhomogeneous} perturbations.
%
%
%
Our goal will be to investigate the influence of the unilateral term $\hat g(v)$ on the Turing instability and on the formation of spatial patterns (spatially nonconstant stationary solutions). From these points of view we will compare the unilateral and classical systems, i.e. a system with a unilateral term $\hat g(v)$ and the corresponding classical system with $\hat g \equiv 0$.


In the classical (smooth) case, we can perform the well known linear analysis
to find necessary conditions for the Turing instability to occur, see e.g. 
\cite{Edelstein1988,JonesSleeman,Murray2003}. 
If these conditions are satisfied then 
starting from small nonhomogeneous disturbances of the ground state,
the solution of \ttg{eq:zdroj}--\ttg{eq:bc1} can converge to another, spatially nonhomogeneous steady state,
provided it exists. In biology, this process of forming nonhomogeneous steady states
can serve as a model of pattern (prepattern) formation mechanisms.
Therefore, we often refer to these spatially nonhomogeneous stationary solutions as patterns.
We will have
$\partial f / \partial u (\bar u, \bar v) > 0$ in the system under consideration
and we will call $u$ the activator. In this case, one of the necessary conditions for the Turing instability
is that the diffusion
coefficient of the activator $u$ is sufficiently smaller than the diffusion 
coefficient of $v$, i.e. the ratio $d_1/d_2$ is sufficiently small.

Our paper is motivated mainly by two surprising results \cite{KucVat2012} and
\cite{SookVat} about systems in the form \ttg{eq:zdroj} with $\hat g \equiv 0$, where $f$ and $g$ satisfy assumptions 
under which Turing instability occurs.
In these papers, a unilateral source is not given by the term $\hat g$ as in \ttg{eq:zdroj}, 
but by certain unilateral conditions for $v$ formulated by variational inequalities.
The former result guarantees existence of 
stationary spatially nonhomogeneous solutions
even for $d_1/d_2$ arbitrarily large. The later result concerns 
certain instability of the ground state for a very wide range of values $d_1$ and $d_2$.
There are also earlier theoretical studies, e.g.\ \cite{DKM,EisKuc,EisKuc1999,EisVat2011,KucBos1994} and references therein,
predicting new and interesting features of systems with various unilateral conditions or terms.
Let us note that the original necessary condition of one fast and one slow diffusion can be removed or relaxed also by other approaches. For example, Turing instability can occur in the presence of cross-diffusion even if the (isolated) activator diffuses faster than the (isolated) inhibitor \cite{Fanelli2013}.
The presence of non-diffusing species yields patterns outside the classical limits of the Turing model \cite{Cantini2014,Klika2012}. 
Both stochastic effects and cross-diffusive terms were shown \cite{Biancalani2010} to yield stochastic self-organization for a wider region of parameters than in the conventional Turing approach.
Similarly, stochastic patterns for a wider region of parameters were observed and analysed in \cite{Asslani2012} for the stochastic Brusselator model. 
Alternatively, nonnormality of the linear stability matrix can amplify the noise and yield fluctuation-induced Turing patterns that are not strongly limited by the value of diffusion coefficients \cite{Biancalani2017}.

Going back to unilateral regulation, the unilateral conditions described by variational inequalities considered 
in \cite{SookVat,KucVat2012}
correspond to sources which
do not allow $v$ to
decrease below zero (i.e. the concentration $V$ is not allowed to decrease 
below the equilibrium concentration $\bar V$) on a given subset of the boundary 
or of the interior of the domain. 
These hard inequalities, however, seem to be unrealistic
from the viewpoint of biological applications, because it is difficult to imagine
a natural mechanism which would strictly prevent the concentration of a morphogen to decrease below the threshold.
Therefore we consider a unilateral term $\hat g(v)$, which seems to be more realistic.
It does not prevent $v$ to decrease below $\theta$, but it works against this decrease. 

Our goal is to investigate this case and 
to find values of the ratio $d_1/d_2$ for which the Turing instability occurs, 
type of the resulting patterns, and possible biological implications. 
These questions have not been addressed before.
Further, we would like to open a question if unilateral sources, which in mathematical models improve conditions for Turing instability and change the resulting form of patterns, may really exist in nature and whether they play a role in spatial patterning observed in biology.


Unilateral terms of the type $\hat g(v)=\tau v^-$ have been introduced in the context of systems 
\ttg{eq:zdroj} under the assumptions guaranteeing the Turing instability already in \cite{EisKuc1999}. 
However,
the stability of the ground state 
has not been analysed
for this type of systems. This analysis is nontrivial, because the possible nonsmoothness 
of the unilateral term  precludes the use of the standard linear analysis.
Moreover, in Section~\ref{se:smooth} below we compare the system with nonsmooth unilateral term and systems, where the nonsmooth term is replaced by smooth approximations. 
Analytical results and numerical computations indicate 
that the ground state in the system with nonsmooth term is stable under different conditions than in systems with
its smooth approximations.
%
%
On the other hand, we also observe that perturbations 
larger than certain minimal size
do evolve to qualitatively similar patterns under the same conditions for both the nonsmooth unilateral term and its smooth approximations. 
Thus, the fact whether a small perturbation of the ground state will evolve to a pattern or not is extremely sensitive to small changes of the nonlinear dynamics near the ground state. A small change of the 
term $\hat g$
in a neighbourhood of zero can turn the stability of the ground state to its instability and vice versa. However, the numerical case study presented in Subsection~\ref{sse:patterns} indicates that if the initial perturbations of the ground state 
are larger than a certain minimal size then they robustly evolve to qualitatively similar patterns regardless small changes of the term $\hat g$ near zero.

Theoretically, it is not clear how to analyse the evolution of perturbations of the ground state that are larger than a certain minimal size. Such theory does not exist. However, the nonsmoothness of the unilateral term could help. We observe, at least in the particular examples presented in Section~\ref{se:smooth}, 
that
%
%
in cases when smooth approximations yield patterns for larger perturbations only, the nonsmooth term yields qualitatively similar patterns even from small perturbations. 
Thus, the question whether larger perturbations of the ground state will evolve to patterns
in systems with (both smooth and nonsmooth) unilateral term or not seems to correspond to
the question of
stability with respect to small perturbations of the system with the nonsmooth unilateral term. 
Theoretical study of the question what are diffusion parameters for which spatially nonhomogeneous stationary solutions exist is done in \cite{EisKuc1999} for the case of nonsmooth terms of the type $\tau v^-$, not for their smooth approximations. Further theoretical results about various other (nonsmooth) unilateral conditions can be found in above mentioned papers.


The rest of this paper is organized as follows.
Section~\ref{se:smooth} shows the significance of the nonsmooth 
unilateral term $\hat g(v) = \tau v^-$ and compares its influence on the initiation and final formation of spatial patterns with the influence of its smooth approximations.
Section~\ref{se:results} presents numerical calculations showing
spatial patterns produced by system \ttg{eq:Maini+g} with $\hat g(v) = \tau v^-$, compares them
with patterns obtained by the same
system without any unilateral term,
and shows how these patterns depend on the strength $\tau$ of the unilateral source
and on the ratio of diffusion constants.
We observe that unilateral terms yield asymmetric patterns with irregular spots. 
Concrete system \ttg{eq:Maini+g} with the unilateral term $\hat g(v) = \tau v^-$ generates patterns even for greater ratio of diffusions in comparison with the classical system.
Finally, we show that 
the difference between the patterns corresponding to the almost zero and high strength 
of the unilateral source resembles the difference between the roughly regular pattern of the 
common morph of the cheetah
and the irregular pattern of the king cheetah.
Section~\ref{se:conslusion} discusses the results and draws the conclusions.

\section{Significance of the nonsmooth unilateral term}
\label{se:smooth}

As we have already mentioned, the unilateral term need not be smooth at the 
point of the ground state and, therefore, the standard linear analysis 
cannot be performed, in general.
If the unilateral term is non-smooth at the ground state, a natural idea is 
to approximate it by a smooth one. Such approximation can be arbitrarily precise 
and therefore we would expect that the behaviour of the approximate system 
will not considerably differ from the behaviour of 
that with the nonsmooth unilateral term.
This vague statement is roughly correct from the perspective of the formation of the final pattern,
but it is not true from the point of view of the Turing instability.
The reason is that the stability is a local effect determined 
by small perturbations of the ground state, but the final pattern 
is formed by nonlinear terms $f$, $g$, and $\hat g$ evaluated at points 
$u$ and $v$ distant from the ground state.
To illustrate this phenomenon,
we provide a short case study to show how various approximations of 
the unilateral term may influence the Turing instability and 
what are their effects on the resulting patterns.
Basically, we show that the occurrence of the Turing instability is extremely sensitive 
on small changes of the nonlinear dynamics near the ground state.

{\bf Particular system.}
We will discuss the particular system used in \cite{BarVarAra1999,LiuLiaMai2006}
for the study of skin and coat patterns in fish and mammals,
and supplement it by a unilateral source term $\hat g(v)$. 
Namely, we will consider the system
\begin{align}
  \label{eq:Maini+g}
  \frac{\rmd u }{\rmd t} &= D\delta \Delta u + \alpha u + v - r_2 u v - \alpha r_3 uv^2
     \quad\text{in }\Omega,\\ \nonumber
  \frac{\rmd v }{\rmd t} &= \delta \Delta v - \alpha u + \beta v  + r_2 u v + \alpha r_3 uv^2 + \hat g(v) \quad\text{in }\Omega.
\end{align}
Note that this system is a special case of \ttg{eq:zdroj} with 
$d_1 = D\delta$, $d_2 = \delta$, 
$f(u,v) = \alpha u + v - r_2 u v - \alpha r_3 uv^2$, and 
$g(u,v) = - \alpha u + \beta v  + r_2 u v + \alpha r_3 uv^2$.
If $\hat g(v) = 0$ then this system coincides with
the original system from \cite{BarVarAra1999,LiuLiaMai2006}
and we call it the classical case.
As in \cite{LiuLiaMai2006}, we will assume the homogeneous Neumann boundary 
conditions \ttg{eq:bc1} and parameter values
\begin{equation}
  \label{eq:param}
   \delta=6,\ \alpha=0.899,\ \beta=-0.91,\ r_2=2,\ r_3=3.5.
\end{equation}
For $D=0.45$ and $\hat g(v)=0$, these values yield the Turing diffusion driven instability \cite{LiuLiaMai2006},
however, we will consider also different values of $D$.

{\bf Ground state.}
The ground state of system \ttg{eq:Maini+g} is defined in the same way
as in Section~\ref{se:intro}, i.e. it consists of constants 
$\bar u, \bar v$ such that 
$f(\bar u, \bar v)=g(\bar u, \bar v)+\hat g(\bar v)=0$.
In particular, it can be readily verified that it is
\begin{equation} 
  \label{eq:baru}
  \bar u = -\bar v / (\alpha - r_2 \bar v - \alpha r_3 \bar v^2),
\end{equation}
where $\bar v$ is a root of the nonlinear equation
\begin{equation}
  \label{eq:barv}
  (1+\beta)\bar v + \hat g(\bar v) = 0.
\end{equation}
Clearly, if $\hat g(0) = 0$ then $\bar u = \bar v = 0$.
This is the case for choices of $\hat g$ we are mainly interested in.
However, certain choices of $\hat g$ introduced below do not vanish at zero
and hence the corresponding ground state is nonzero.

{\bf Conditions for the Turing instability.}
In the case when the additional unilateral term $\hat g(v)$ in \ttg{eq:Maini+g} is smooth at $\bar v$,
we can perform the standard linear analysis to obtain necessary conditions 
for the Turing instability, see e.g. \cite{Edelstein1988,JonesSleeman,Murray2003}.
Namely, we can introduce the Jacobi matrix of the map $f$, $g + \hat g$ at $\bar u$, $\bar v$ as
\begin{equation}
  \label{eq:defB}
  B = \left[ \begin{array}{cc} b_{11}, & b_{12} \\ b_{21}, & b_{22} \end{array} \right]
    = \left[ \begin{array}{cc} \partial f / \partial u, & \partial f / \partial v \\
     \partial g / \partial u, & \partial g / \partial v +  \rmd \hat g / \rmd v \end{array} \right]
     (\bar u, \bar v).
\end{equation}
If 
\begin{equation}
  \label{eq:trBdetB}
  \operatorname{tr}  B < 0 
  \quad\text{and}\quad
  \det B > 0,
\end{equation}
then 
the ground state $(\bar u,\bar v)$ is asymptotically stable with respect to small spatially homogeneous perturbations.
If simultaneously
\begin{equation}
  \label{eq:bij}
b_{11} b_{22} < 0 \mbox{ and } b_{12}b_{21} < 0
\end{equation}
then this ground state is stable 
(with respect to small spatially nonhomogeneous perturbations) 
only for some values of $D$
and unstable for the others,
see e.g. \cite[sec.~2.3]{Murray2003}.

{\bf Critical ratio of diffusions.}
Parameter values \ttg{eq:param} are chosen in such a way that for $\hat g \equiv 0$ conditions \ttg{eq:trBdetB} and \ttg{eq:bij} are fulfilled. 
In any case, if $\hat g(v)$ is smooth at $\bar v$ and if conditions \ttg{eq:trBdetB} and \ttg{eq:bij} hold then
a necessary condition for the ground state of system \ttg{eq:Maini+g} to be unstable with respect to spatially nonhomogeneous perturbations is that the ratio of diffusion coefficients $D$ is sufficiently small. Precisely, the condition is
\begin{equation}
  \label{eq:dcrit}
D < \dcrit 
\quad\text{with}\quad
 \dcrit = \frac{1}{b_{22}^2} \left(
    \det B - b_{12}b_{21} - 2\sqrt{-b_{12}b_{21}\det B}
  \right).
\end{equation}
Note that the definition of $\dcrit$ in \ttg{eq:dcrit} is just a reciprocal value of the formula from 
\cite[p.~562]{Nishiura1982}. It can also be easily derived from the analysis of
\cite[p.~109]{Murray2003}.
In any case, if condition \ttg{eq:dcrit} is not satisfied then 
the Turing instability cannot occur.
It is essential that if $\hat g$ is not smooth at the ground state value $\bar v$ then 
this linear analysis cannot be performed. Jacobian $B$ is simply not defined 
and consequently formula \ttg{eq:dcrit} has no sense.  
The critical ratio $\dcrit$ can be only estimated numerically. 

The forthcoming Subsection~\ref{sse:nodiffusion} discusses the stability of the ground state with respect to spatially homogeneous perturbations, i.e. the stability for the system without diffusion.
Subsection~\ref{sse:instability} defines six particular choices of $\hat g(v)$ 
and compares them with respect to the Turing instability.
We emphasize that the Turing
instability is a local phenomenon determined by \emph{small} perturbations and hence only 
the values of $\hat g(v)$ in a small neighbourhood of the ground state $\bar v$ are relevant.  
We will see that although the six choices of $\hat g(v)$ differ only slightly in the neighbourhood of the ground state, some of them yield patterns evolving from small
perturbations and some of them do not. 
However, further in Subsection~\ref{sse:patterns} we will see that if the perturbations of the ground state are \emph{larger} than a certain minimal size, 
then they evolve to qualitatively similar patterns in all cases.

We note that all patterns in this paper are computed numerically by our own Matlab based finite element 
solver. The convergence and stability of the finite element method is well known \cite{Ciarlet1978}. Its convergence for the specific problem \eqref{eq:Maini+g} is analysed in \cite{Kus2015}.
The time stepping is done by the build in Matlab ODE solver, which stops as soon as the prescribed final time is reached. We set this time to $5\cdot10^4$ for all presented calculations. This is a sufficiently high value, because the experimentally determined times of reaching the steady state are usually around $2$--$3\cdot10^3$, at most $10^4$. The high value of the final time does not increase the total computational time considerably, because the method determines the time step adaptively. As soon as the numerical method detects the steady state, the time step quickly increases and the final time is reached in a few iterations.
The initial condition is always chosen as small random fluctuations around the ground state, except of Figure~\ref{fi:patternsdef}, where the fluctuations are larger. Clearly, different initial conditions may and often do evolve to different stationary solutions, but qualitative features of these solutions are the same.
We choose the domain to be $\Omega = (-100,100)^2$ and
in the subsequent figures, we plot the patterns as graphs of the solution component $u$, where values of $u$ are indicated by shades of grey. We do not plot the component $v$, because it is complementary to $u$, cf. \cite[p.~88]{Murray2003}, and patterns based on $v$ are almost exact inverses of patterns based on $u$.

\subsection{{\bf Stability for systems without diffusion}}
\label{sse:nodiffusion}

In this subsection, we consider system \ttg{eq:Maini+g} without the diffusion terms and we analyse the stability of its ground state. We are mainly interested in the case $\hat g(v) = \tau v^-$, i.e. in the natural choice $\theta = 0$, and study
the system
\begin{align} \label{eq:ODE+tauv-}
  \frac{\rmd u }{\rmd t} &= \alpha u + v - r_2 u v - \alpha r_3 uv^2,
\\ \nonumber
  \frac{\rmd v }{\rmd t} &= - \alpha u + \beta v  + r_2 u v + \alpha r_3 uv^2 + \tau v^-.
\end{align}
The ground state of this system is $(\bar u, \bar v) = (0,0)$ and its stability cannot be investigated by the standard linearisation due to the nonsmoothness of the unilateral term. Nevertheless, we can prove it using, for example, the theory of Filippov systems \cite{Filippov:1988}.

\begin{lemma}\label{le:stabODE}
If $\alpha$, $\beta$ are given by \ttg{eq:param} and $\tau \in (0, 0.09)$ then the trivial solution $(\bar u, \bar v) = (0,0)$ of system \ttg{eq:ODE+tauv-} 
is asymptotically stable.
\end{lemma}
\begin{proof}
Let us start with the system without higher order terms, i.e.
\begin{align}
  \label{eq:ODElin+g}
  \frac{\rmd u }{\rmd t} 
  = \alpha u + v, \ \
  \frac{\rmd v }{\rmd t} 
  = - \alpha u + \beta v + \tau v^-.
\end{align}
Let $U(t)=(u(t),v(t))$ be its arbitrary solution. 
If $\alpha$, $\beta$ are given by \ttg{eq:param} and $t_0$ is such that 
$v(t_0)=0$ and $u(t_0)>0$, then it follows from \ttg{eq:ODElin+g} that 
$\frac{\rmd u }{\rmd t}(t_0)>0$ and $\frac{\rmd v }{\rmd t}(t_0)<0$.
Similarly, if  $v(t_0)=0$ and $u(t_0)<0$ then
$\frac{\rmd u }{\rmd t}(t_0)<0$ and $\frac{\rmd v }{\rmd t}(t_0)>0$.
Thus, the solution intersects the axis $v=0$ transversally and the whole time interval $(0,+\infty)$
consists of open intervals where $v>0$, open intervals where $v<0$, and isolated points. 
In time intervals where $v(t)>0$, the solution $U(t)$ 
coincides with a solution of the linear system obtained by the choice $\tau =0$,
and in time intervals where $v(t)<0$, it coincides with a solution of the linear system
\begin{align}
  \label{eq:ODElin+tau}
  \frac{\rmd u }{\rmd t} = \alpha u + v, \ \ 
  \frac{\rmd v }{\rmd t} = - \alpha u + \beta v - \tau v.
\end{align}
For $\alpha$, $\beta$ given by \ttg{eq:param}, 
the matrix $B_0$ of system \ttg{eq:ODElin+g} with $\tau =0$ satisfies \ttg{eq:trBdetB}.
The trace of the matrix $B_\tau$ of system \ttg{eq:ODElin+tau} 
is negative for all $\tau >-0.011$ and its determinant
is positive for $\tau <0.09$. It follows that for $\tau \in (0, 0.09)$, 
the eigenvalues of both matrices $B_0$ and $B_\tau$ have negative real parts, that means the trivial solution of both linear systems is asymptotically stable. 
Due to the form of solutions of linear systems with constant coefficients, 
there exists $C>0$ such that if $U(t)$ is any solution of \ttg{eq:ODElin+g} with $\tau =0$ or \ttg{eq:ODElin+tau} then
$$
\|U(t)\|\le C\|U(0)\| \exp(\Lambda t),  
$$
where $\Lambda$ is the maximum over real parts of all eigenvalues of both matrices 
$B_0$ and $B_\tau$. Using considerations above we see that also the trivial solution of
\ttg{eq:ODElin+g} is asymptotically stable.
Further, it follows from known results concerning systems with nonsmooth right hand sides that then the trivial solution of the full system 
\ttg{eq:ODE+tauv-} with higher order terms is also asymptotically stable, see e.g. \cite[p.~169, Theorem~7]{Filippov:1988}.
\end{proof}

We note that in fact the trivial solution of \ttg{eq:ODE+tauv-} is stable for a larger interval of $\tau$ 
because its stability for both linear problems mentioned is only sufficient, not necessary condition.
Further we note that in all situations discussed below we will have $\tau < 0.09$ and therefore the trivial 
solution of our problem without any diffusion will be always stable.

As we already mentioned, we will also investigate the effect of smoothing of the nonsmooth unilateral term $\tau v^-$. We will consider several approximations in the form $\hat g(v) = \tau n_\varepsilon(v)$, where the parameter $\varepsilon > 0$ controls the accuracy of the approximation.
Functions $n_\varepsilon(v)$ are smooth in a neighbourhood of the ground state and approximate $v^-$ in the sense that $n_\varepsilon(v)$ converge to $v^-$ as $\varepsilon\to0_+$.
This setting enables to analyse the stability of the ground state for all approximations of this type at once.

Thus, we consider the ODE system
\begin{align}
  \label{eq:ODE+n}
  \frac{\rmd u }{\rmd t} &= \alpha u + v - r_2 u v - \alpha r_3 uv^2,
     \\ \nonumber
  \frac{\rmd v }{\rmd t} &= - \alpha u + \beta v  + r_2 u v + \alpha r_3 uv^2 + \tau n_\varepsilon (v),
\end{align}
where the parameter values are given in \ttg{eq:param} and $\tau\geq 0$, $\varepsilon > 0$ are free parameters.
We assume that \ttg{eq:barv} with $\hat g(v) = \tau n_\varepsilon(v)$ has a unique solution $\bar v_\varepsilon$ for any $\varepsilon > 0$. This $\bar v_\varepsilon$ together with $\bar u_\varepsilon$ given by \ttg{eq:baru} forms the ground state of \ttg{eq:ODE+n}.

\begin{lemma}\label{le:g_smooth}
Let $(\bar u_\varepsilon,\bar v_\varepsilon)$ be the ground state of system \ttg{eq:ODE+n}. 
For any $\varepsilon >0$, let the function $n_\varepsilon$ be smooth in a neighbourhood of $\bar v_\varepsilon$ and let 
$n_\varepsilon'(v)$ denote the derivative of $n_\varepsilon(v)$. 
If $n_\varepsilon(v)$ converges uniformly in $\R$ to $v^-$ as $\varepsilon\to 0_+$
and if 
\begin{equation}
\label{eq:tilden}
\tilde n := \lim_{\varepsilon \to 0_+} n_\varepsilon'(\bar v_\varepsilon) \text{ exists,}\quad
\tilde n \leq 0, \quad \text{and}\quad
0 \leq \tau < (1+\beta) \min\{1, -1/\tilde n \},
\end{equation}
then 
the ground state $(\bar u_\varepsilon,\bar v_\varepsilon)$ of system \ttg{eq:ODE+n}
is asymptotically stable for all sufficiently small $\varepsilon>0$.
\end{lemma}

\begin{proof}
First, we notice that the ground state tends to zero as $\varepsilon \to 0_+$. This clearly follows from \ttg{eq:baru}  and the convergence $\bar v_\varepsilon \to 0$ which can be proved by contradiction.
Indeed, if $\bar v_\varepsilon$ did not converge to $0$ then there would be a sequence $\varepsilon_k \to 0$ such that $\bar v_{\varepsilon_k} \to v_0$ with $v_0 \ne 0$.
Due to \ttg{eq:barv}, we would have the identity 
\begin{equation}
  \label{eq:vepsk}
  0 = (1+\beta) \bar v_{\varepsilon_k} + \tau n_{\varepsilon_k} (\bar v_{\varepsilon_k}).
\end{equation}  
If $v_0$ were finite then using the uniform convergence $n_\varepsilon(v) \to v^-$ 
we would get $(1+\beta) v_0 + \tau v_0^- = 0$. However, this equation has only the zero solution because $\tau < 1+\beta$.
This would contradict the assumption $v_0 \neq 0$. It is not hard to see, again by using the uniform convergence and the assumption 
$\tau < 1+\beta$, that if $v_0$ were not finite then \ttg{eq:vepsk} could not hold for large $k$.  


Second, we show the asymptotic stability.
We linearise system \ttg{eq:ODE+n} around $(\bar u_\varepsilon, \bar v_\varepsilon)$. 
The matrix $B_\varepsilon$ of this linearisation is given by \ttg{eq:defB} with
$f(u,v) = \alpha u + v - r_2 u v - \alpha r_3 uv^2$, $g(u,v) = - \alpha u + \beta v  + r_2 u v + \alpha r_3 uv^2 + \hat g(v)$, $\hat g(v) = \tau n_\varepsilon (v)$, and $(\bar u, \bar v) = (\bar u_\varepsilon, \bar v_\varepsilon)$.
Due to convergences $(\bar u_\varepsilon, \bar v_\varepsilon) \to (0,0)$ and $n_{\varepsilon}'(\bar v_\varepsilon) \to \tilde n$, we have 
$$
  \lim_{\varepsilon \to 0_+} B_\varepsilon = \left[ \begin{array}{cc}
    \alpha, & 1 \\
    -\alpha, & \beta + \tau \tilde n
  \end{array}\right].
$$
Under the assumption $0 \leq \tau < -(1+\beta)/\tilde n$, we can easily verify that this matrix satisfies conditions \ttg{eq:trBdetB}. Thus, the continuity argument implies that even the matrix $B_\varepsilon$ satisfies conditions \ttg{eq:trBdetB} for sufficiently small $\varepsilon > 0$ and, consequently, the stationary solution $(\bar u_\varepsilon, \bar v_\varepsilon)$ is asymptotically stable. 
\end{proof}

\subsection{{\bf Turing instability for various choices of $\hat g(v)$}}
\label{sse:instability}

Now, we will consider several particular choices of $\hat g(v)$ 
and compare their influence on the Turing instability,
i.e. on the evolution of \emph{small} spatially nonhomogeneous perturbations of the ground state.
The idea is to consider the unilateral source term $\hat g(v) = \tau v^-$ as the reference choice
and the other cases are seen as its approximations.
As a criterion for the comparison we choose  
the critical ratio $\dcrit$, see \ttg{eq:dcrit} and the discussion below \ttg{eq:dcrit}. We will see that $\dcrit$ varies considerably for different choices of $\hat g(v)$ and that this variation is essential even in the case of very accurate approximations. 
Note that the strength of the unilateral source is 
\begin{equation}
\label{eq:tauval}
  \tau = 0.075
\end{equation}
for all cases throughout this section.
Now, we list the choices of $\hat g(v)$ we make.


\begin{figure} 
%
%
%
\parbox{0.3\textwidth}{
\raisebox{0.19\textwidth}{(a)}
\includegraphics[width=0.25\textwidth]{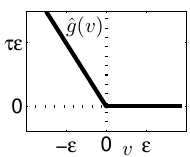}
}
\parbox{0.3\textwidth}{
\raisebox{0.19\textwidth}{(b)}
\includegraphics[width=0.25\textwidth]{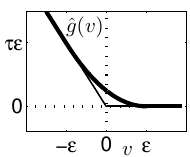}
}
\parbox{0.3\textwidth}{
\raisebox{0.19\textwidth}{(c)}
\includegraphics[width=0.25\textwidth]{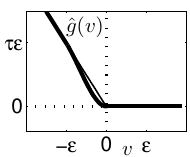}
}
\\[12pt]
\parbox{0.3\textwidth}{
\raisebox{0.19\textwidth}{(d)}
\includegraphics[width=0.25\textwidth]{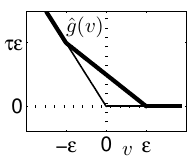}
}
\parbox{0.3\textwidth}{
\raisebox{0.19\textwidth}{(e)}
\includegraphics[width=0.25\textwidth]{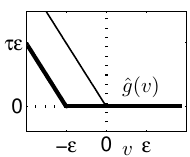}
}
\parbox{0.3\textwidth}{
\raisebox{0.19\textwidth}{(f)}
\includegraphics[width=0.25\textwidth]{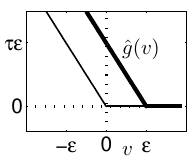}
}


\caption{\label{fi:choices}
Graphs of different choices of the unilateral term $\hat g(v)$ in \ttg{eq:Maini+g}: 
(a) $\hat g(v) = \tau v^-$,
(b) $\hat g(v)$ given by \ttg{eq:smooth},
(c) $\hat g(v)$ given by \ttg{eq:smooth2},
(d) $\hat g(v)$ given by \ttg{eq:lincut},
(e) $\hat g(v) = \tau(v+\varepsilon)^-$,
(f) $\hat g(v) = \tau(v-\varepsilon)^-$.
Thin lines show the graph of $\tau v^-$ for comparison.
}

\end{figure}

\begin{figure} 
\includegraphics[width=0.75\textwidth]{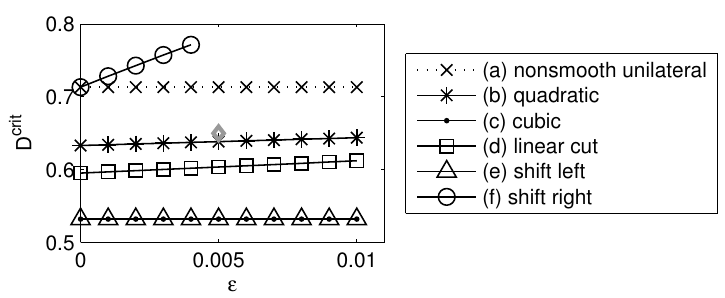} 
\caption{\label{fi:Dcrit}
Dependence of the critical ratio of diffusions $\dcrit$ on $\varepsilon$ for choices (a)--(f). The value for choice (a) is estimated numerically and the other values are computed by \ttg{eq:dcrit}. 
Notice that cases (c) and (e) coincide. In case (f), the critical ratio is not defined for $\varepsilon$ above approximately $0.0044$, because then the trace of $B$ is positive, see \ttg{eq:trBdetB}.
The grey diamond indicates the case of parameter values $\varepsilon=0.005$ and $D=0.65$ for which Turing patterns appear in the case (a) only.
}
\end{figure}

\begin{description}

\item[(a) Nonsmooth unilateral source, $\hat g(v) = \tau v^-$, see Figure~\ref{fi:choices}(a).]
This is the reference case.
The ground state of system \ttg{eq:Maini+g} with this choice of $\hat g(v)$ is $(\bar u, \bar v) = (0,0)$.
Its asymptotic stability with respect to small spatially homogeneous perturbations, i.e. as a stationary solution of ODE system \ttg{eq:ODE+tauv-}, follows from Lemma~\ref{le:stabODE}.
The instability of this ground state as a solution of \ttg{eq:Maini+g} cannot be investigated by the linear analysis,
but numerical calculations indicate that spatially nonhomogeneous perturbations as small as we can afford numerically, evolve to nonhomogeneous stationary solutions for the ratio of diffusions below $\dcrit[(a)]=0.71$. Note that this value is greater than the critical ratio of diffusion for the classical case ($\hat g \equiv 0$), which is $\dcrit=0.53$. See Section~\ref{se:results} for more details. 



\item[(b) Smooth quadratic approximation.]
The nonsmooth function from the previous case can be smoothed
for example as
\begin{equation}
\label{eq:smooth}
  \hat g(v) = \left\{ \begin{array}{cc}
    \tau (v-\varepsilon)^2/(4\varepsilon) & \text{for } |v| < \varepsilon, \\
    \tau v^- & \text{for } |v| \geq \varepsilon,
  \end{array} \right. 
\end{equation}
where $\varepsilon > 0$ is a small parameter, see Figure~\ref{fi:choices}(b).
System \ttg{eq:Maini+g} with \ttg{eq:bc1} and this choice of $\hat g(v)$
has the ground state with $\bar u_\varepsilon$ given by \ttg{eq:baru} and 
$$
  \bar v_\varepsilon = 2 \frac{\varepsilon}{\tau} \left( -1-\beta+\frac{\tau}{2}+\sqrt{(1+\beta)(1+\beta-\tau)} \right).
$$
For the chosen parameters, assumptions \ttg{eq:tilden} in Lemma~\ref{le:g_smooth} are satisfied, because, in particular, $\tilde n \approx -0.71$.
%
%
Thus, the asymptotic stability of $(\bar u_\varepsilon, \bar v_\varepsilon)$ of \ttg{eq:ODE+n} follows
for sufficiently small $\varepsilon > 0$. Moreover, numerically we can easily find that it is asymptotically stable for $\varepsilon < 0.038$.
The corresponding critical ratio of diffusions can be expressed from \ttg{eq:dcrit} for any $0 < \varepsilon < 0.038$ and we plot its values in Figure~\ref{fi:Dcrit}. In the limit $\varepsilon \rightarrow 0_+$ we have $\dcrit[(b)] \approx 0.63$.


\item[(c) Smooth cubic approximation.]
Another option how to smooth the function from the case (a) is
\begin{equation}
\label{eq:smooth2}
  \hat g(v) = \left\{ \begin{array}{cl}
    \tau v^2 (v+2\varepsilon)/\varepsilon^2 & \text{ for } -\varepsilon < v < 0, \\
    \tau v^- & \text{ for } v \leq -\varepsilon \text{ and } v \geq 0,
  \end{array} \right.
\end{equation}
where $\varepsilon > 0$ is again a small parameter, see Figure~\ref{fi:choices}(c).
The ground state in this case is $(\bar u, \bar v) = (0,0)$ independently of $\varepsilon$ and since the derivative of $\hat g$ at zero vanishes, the critical ratio of diffusions is the same as in the classical system ($\hat g \equiv 0$). Using \ttg{eq:dcrit} we obtain $\dcrit[(c)]=\dcrit \approx 0.53$. 

\item[(d) Linear cut.]
The choice (a) can be approximated by a continuous piecewise linear 
function such that it is smooth at the ground state. A straightforward choice is
\begin{equation}
\label{eq:lincut}
  \hat g(v) = \left\{ \begin{array}{cc}
    \tau (\varepsilon-v)/2 & \text{for } |v| < \varepsilon, \\
    \tau v^- & \text{for } |v| \geq \varepsilon,
  \end{array} \right. 
\end{equation}
see Figure~\ref{fi:choices}(d).
The ground state is shifted away from zero. Its component $\bar u_\varepsilon$ is given by \ttg{eq:baru} and
$$
  \bar v_\varepsilon =  \frac{\tau\varepsilon}{\tau-2-2\beta}.
$$
Using Lemma~\ref{le:g_smooth} for parameter values \ttg{eq:param} and \ttg{eq:tauval}, we obtain the asymptotic stability of the ground state $(\bar u_\varepsilon,\bar v_\varepsilon)$ of \ttg{eq:ODE+n} for sufficiently small $\varepsilon > 0$, because $\tilde n = -1/2$ and assumptions \ttg{eq:tilden} are satisfied.
Numerically, we verify that it is asymptotically stable for $\varepsilon \leq 0.016$.
The critical ratio of diffusions is computed from \ttg{eq:dcrit} and its dependence on $\varepsilon$ 
is illustrated in Figure~\ref{fi:Dcrit}.
The limit value for $\varepsilon \to 0_+$ is $\dcrit[(d)] \approx 0.60$.

\item[(e) Shift of the threshold to the left, $\hat g(v) = \tau (v + \varepsilon)^-$, see Figure~\ref{fi:choices}(e).]
The corresponding ground state is $(\bar u, \bar v) = (0,0)$ for all $\varepsilon > 0$. This $\hat g(v)$ is smooth at zero and both its value and derivative at zero vanish. Therefore, conditions for the Turing instability 
are the same as in the classical case $\hat g \equiv 0$ and
formula \ttg{eq:dcrit} provides the same critical ratio of diffusions as in the case (c), i.e. $\dcrit[(e)] = \dcrit[(c)] = \dcrit \approx 0.53$.

\item[(f) Shift of the threshold to the right, $\hat g(v) = \tau (v - \varepsilon)^-$, see Figure~\ref{fi:choices}(f).] 
This choice yields a nonzero ground state $\bar u_\varepsilon$ given by \ttg{eq:baru} and 
$$
  \bar v_\varepsilon = \frac{\tau\varepsilon}{\tau-1-\beta}.
$$
Using Lemma~\ref{le:g_smooth} for parameters \ttg{eq:param} and \ttg{eq:tauval}, we obtain the asymptotic stability of the ground state $(\bar u_\varepsilon, \bar v_\varepsilon)$ of \ttg{eq:ODE+n} for sufficiently small $\varepsilon > 0$,
because $\tilde n = -1$ and assumptions \ttg{eq:tilden} are satisfied. 
Numerically, we verify that it is asymptotically stable for $\varepsilon < 0.0044$.
The critical ratio of diffusions can be obtained from \ttg{eq:dcrit}. Its values are provided in Figure~\ref{fi:Dcrit} and its limit for $\varepsilon \rightarrow 0_+$ is $\dcrit[(f)] \approx 0.71$.

\end{description}

To compare choices (a)--(f), we summarize the dependence of the critical ratio of diffusions on $\varepsilon$ in Figure~\ref{fi:Dcrit}. Note that the accuracy of approximations (b)--(f) of the reference choice (a) is controlled by $\varepsilon$. Smaller $\varepsilon$ corresponds to more accurate approximations.
We clearly observe that different choices of $\hat g(v)$ yield considerably different critical ratios of diffusions and, hence, various approximations of the nonsmooth unilateral term $\tau v^-$ exhibit the Turing instability for
different values of the ratio $D$.
For example,
if $\varepsilon = 0.005$ and $D = 0.65$ (see the grey diamond in Figure~\ref{fi:Dcrit}),
then choice (a) is the only case which exhibits the Turing instability. 
Indeed, in cases (b)--(e)
the ground state is stable with respect to all small perturbations, because the ratio of diffusion $D=0.65$ is above the critical value. 
In the case (f) the Turing instability cannot occur, because $\mathrm{tr}\,B$ is positive for 
$\varepsilon > 0.0044$ 
and therefore the ground state is unstable even with respect to spatially homogeneous perturbations.
Similarly, if we decrease $\varepsilon$ to $0.001$ and keep $D = 0.65$, then
the choices (a) and (f) exhibit the Turing instability, but choices (b)--(e) do not.

Note that our statement that the choice (a) exhibits the Turing instability is based on numerical calculations, where we observe that small perturbations of the ground state evolve to patterns. For more details and numerical results,
see Figure~\ref{fi:vzornik} below, the panels for $\tau = 0.075$.


Figure~\ref{fi:Dcrit} clearly shows the size of variations in $\dcrit$
for different approximations of the nonsmooth unilateral source term. 
Even if we arbitrarily increase the accuracy of these approximations, i.e. in the limit $\varepsilon \rightarrow 0$, the corresponding values of $\dcrit$ differ considerably. 
Hence, each approximation yields the Turing instability for different ranges of diffusion coefficients.
Consequently, 
\emph{the idea to approximate the nonsmooth term} by a smooth one and analyse it by standard means \emph{fails}.
Simple smooth approximations of the nonsmooth unilateral term with accuracy controlled by $\varepsilon$ can yield misleading results in the limit $\varepsilon \rightarrow 0$.

From the biological perspective, all choices (a)--(f) seem to be plausible.
Although some of these choices have the threshold value shifted from the ground state,
the difference is not large.

Of course, we could also discuss other approximations of the nonsmooth unilateral term.
However, results of the next subsection indicate that these approximations result to the same (or very similar) patterns as the unilateral term $\hat g(v) = \tau v^-$,
provided the other parameters of the problem are the same.
At the same time, it is important to mention
that not all of these approximations yield patterns developing from small perturbations. Sometimes, the perturbations have to be sufficiently distant from the ground state, as we describe below.

\subsection{{\bf Patterns for various choices of $\hat g(v)$}}
\label{sse:patterns}

\begin{figure} 
\makebox[0pt][l]{}\hspace{-1.5mm}%
\makebox[0pt][l]{\raisebox{0.300\textwidth}{(a)}}%
\includegraphics[height=0.29\textwidth]{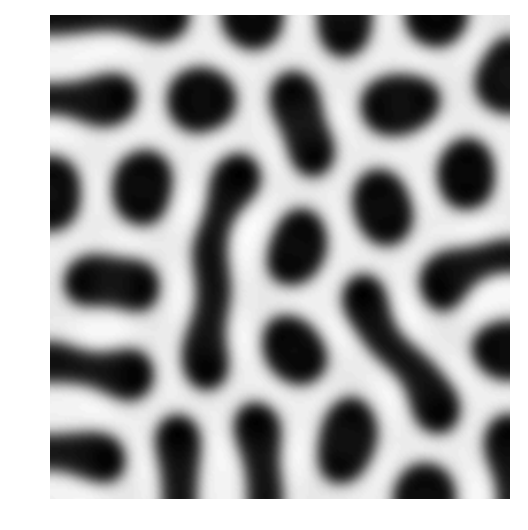}
\makebox[0pt][l]{\raisebox{0.300\textwidth}{(b)}}%
\includegraphics[height=0.29\textwidth]{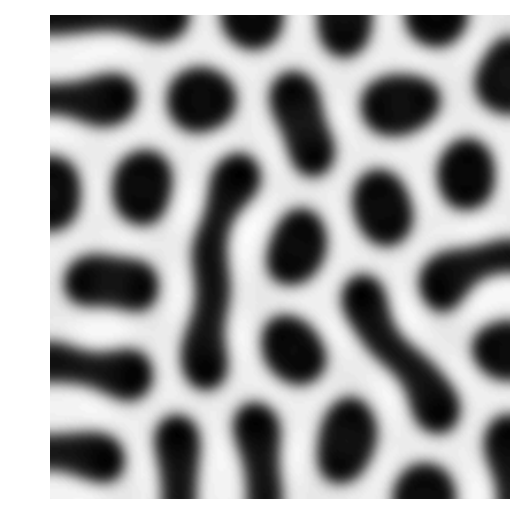}
\makebox[0pt][l]{\raisebox{0.300\textwidth}{(c)}}%
\includegraphics[height=0.29\textwidth]{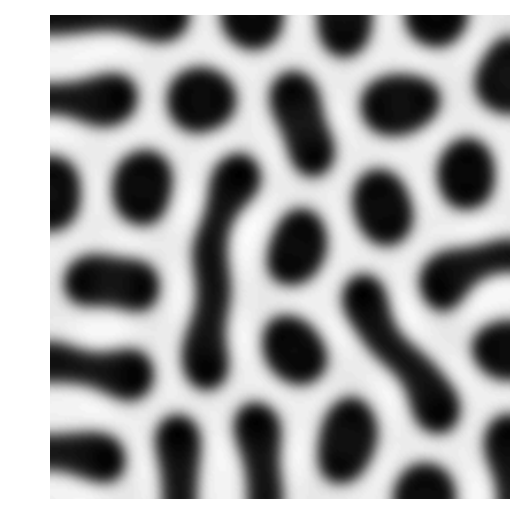}
\makebox[0.11\textwidth][l]{}
\\[6pt]
\makebox[0pt][l]{}\hspace{-1.5mm}%
\makebox[0pt][l]{\raisebox{0.295\textwidth}{(d)}}
\includegraphics[height=0.29\textwidth]{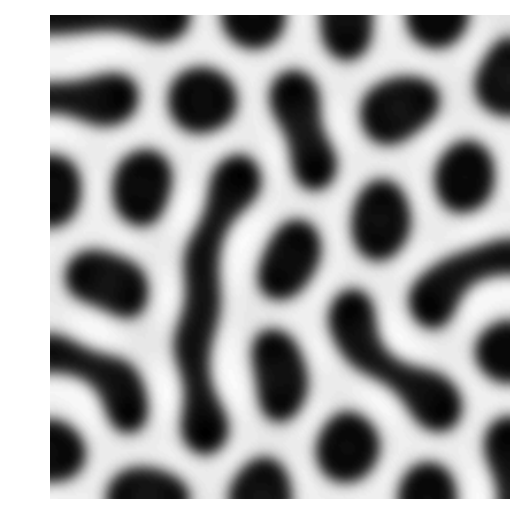}
\makebox[0pt][l]{\raisebox{0.295\textwidth}{(e)}}%
\includegraphics[height=0.29\textwidth]{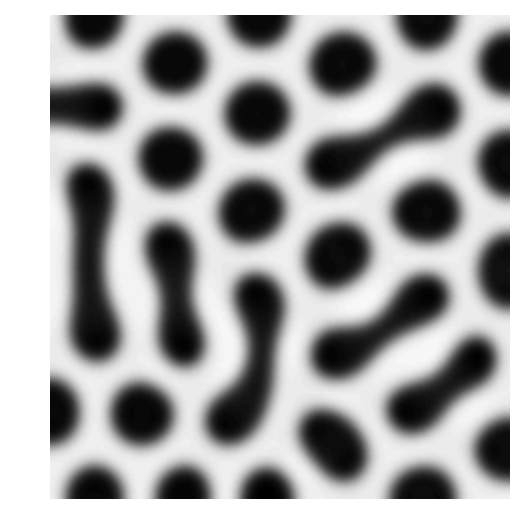} 
\makebox[0pt][l]{\raisebox{0.295\textwidth}{(f)}}%
\includegraphics[height=0.29\textwidth]{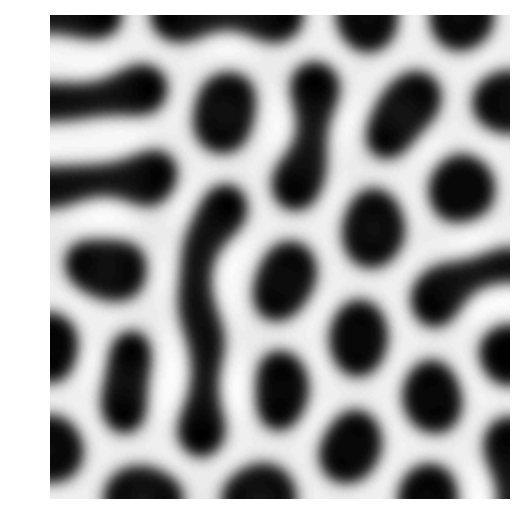}%
\makebox[0.11\textwidth][r]{\raisebox{0pt}[0pt][0pt]{\includegraphics[height=0.57\textwidth]{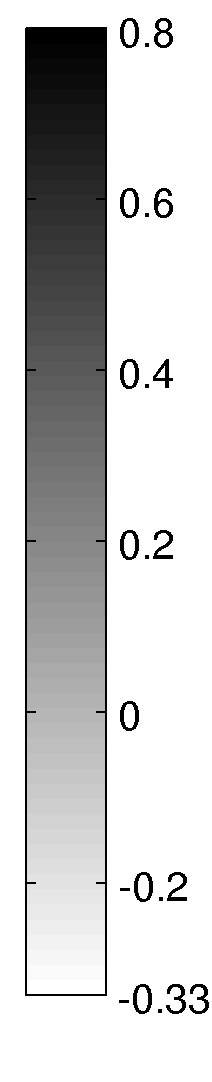}}}%
\caption{\label{fi:patternsdef}
Patterns produced from larger perturbations of the ground state in cases (a)--(f) with $D=0.65$, $\varepsilon=0.005$, $\tau=0.075$,
and parameter values \ttg{eq:param}. 
The initial condition is the same in all cases and consists of
random disturbances of the ground state with maximal amplitude $20\varepsilon=0.1$.
Note that initial conditions with maximal amplitude of size $\varepsilon$ or smaller evolve to patterns in case (a) only.
}
\end{figure}

Above, we introduced an example of parameter values for system \ttg{eq:Maini+g}
such that \emph{small} perturbations of the ground state do not evolve to any patterns in cases (b)--(f), but they do in the case (a), see the grey diamond in Figure~\ref{fi:Dcrit}.
Now, we will see that perturbations \emph{larger} than a certain minimal size
(depending on $\varepsilon$) do evolve to patterns in all these cases.
We also show that all these patterns are qualitatively the same. Moreover, if they evolve from the same initial condition, they are all also quantitatively very similar and some of them are even exactly identical, see Figure~\ref{fi:patternsdef}.

Similarity of these patterns is not surprising, because the differences among all choices of $\hat g(v)$ in cases (a)--(f) are insignificant on scales considerably larger than $\varepsilon$. 
Since the magnitude of the final pattern (i.e. the stationary solution to \ttg{eq:Maini+g}) is of order one and the size of $\varepsilon$ is of order one thousands, we can expect similar patterns in all these cases. 

As we have mentioned, small perturbations of the ground state do not evolve to any patterns in cases (b)--(f) for the chosen values $D=0.65$ and $\varepsilon = 0.005$. This fact follows from the linear analysis and we observe it numerically as well. However, Figure~\ref{fi:patternsdef} shows that perturbations that are larger than a certain minimal size, do evolve to patterns in all these cases. Moreover, we observe identical patterns for choices (a)--(c)
and a very similar pattern for the choice (d). Choices (e) and (f) yield slightly more distinct patterns, but they share the same qualitative features as the other cases.

We also computed the patterns starting from an initial condition twice as large as was used in Figure~\ref{fi:patternsdef} and we obtained identical patterns for all cases (a)--(d). (These results are not presented.) Patterns (e) and (f) were different in a similar manner as in Figure~\ref{fi:patternsdef}.
This is understandable, because choices (a)--(d) of $\hat g(v)$ are identical for $|v| \geq \varepsilon$ and thus if the size of the initial condition is sufficiently large, the influence of $\hat g(v)$ for $|v| \geq \varepsilon$ overweights the influence of $\hat g(v)$ for $|v| < \varepsilon$ and identical patterns emerge.
On the contrary, in cases (e) and (f) the values $\hat g(v)$ slightly differ even for $|v| \geq \varepsilon$ and therefore the resulting patterns differ as well.
Hence, for the particular system \eqref{eq:Maini+g} it seems that
if two nonlinear kinetics differ on a small neighbourhood of the (unique) ground state only then
there exists (almost) the same pattern for both kinetics and
sufficiently large initial perturbations of the ground state will evolve to this pattern for both kinetics.
If this statement is true then practically relevant is the evolution of perturbations greater than a certain minimal size
rather than the evolution of small perturbations. 
The reason is the robustness of the evolution of
the larger perturbations to patterns observed in numerical tests described in this section and the fact that the stability with respect to the small perturbations is highly sensitive to small changes of $\hat g(v)$ in the neighbourhood of the ground state.

Importantly, the nonsmooth unilateral term (a) yields patterns that evolve from small spatial perturbations for a large range of values of $D$, as far as we can conclude from numerous numerical calculations we performed. 
This is the essential motivation to investigate the nonsmooth unilateral case (a). 
It provides predictions about a whole class of approximations
of the nonsmooth term $\hat g(v) = \tau v^-$. The tested choices (b)--(f) are just examples of members of this class. All approximations from this class produce the desired patterns and all these patterns are similar, however, for certain approximations the patterns do not evolve from small perturbations.
There is no known theory so far that would explain
the evolution of initial perturbations that are not small.
However, the approaches presented in \cite{EisKuc1999,SookVat,KucVat2012} provide certain ideas how
to treat theoretically the positive homogeneous nonsmooth case $\hat g(v) = \tau v^-$.
And, as we have already mentioned, the stability and instability of the ground state in systems with this term seem to correspond to the question whether the larger perturbations of the ground state do evolve to patterns or not for systems where this term is approximated.

\section{
Existence of patterns and their dependence on parameters
}
\label{se:results}


In this section we further investigate system \ttg{eq:Maini+g}
with the nonsmooth unilateral term $\hat g(v) = \tau v^-$
to show when the Turing instability occurs, 
what is the effect of this term on the resulting patterns,
and how they depend on the strength $\tau$ and on
the ratio of diffusions $D$. 
In addition, we numerically compare behaviour of the system with this nonsmooth unilateral term
with cases $\hat g \equiv 0$ and $\hat g(v) = -\tau v$. 
Comparing to these linear choices of $\hat g(v)$, we show that the unilateral term produces irregular patterns.
Further, we present numerical results indicating that 
system \ttg{eq:Maini+g} with the nonsmooth term $\hat g(v) = \tau v^-$ yields patterns for considerably higher
ratio of diffusion constants comparing to the classical system with $\hat g \equiv 0$. 
In addition, the choice $\hat g(v) = -\tau v$ seems to be informative about the Turing instability
of the nonsmooth unilateral term.
For system \ttg{eq:Maini+g}, we present results of numerical calculations supporting the hypothesis that
the Turing instability occurs in the nonsmooth unilateral case $\hat g(v) = \tau v^-$
for the same ratio of diffusion coefficients as for the choice $\hat g(v) = -\tau v$.
Finally, in the last part of this section, we compare the patterns obtained with the nonsmooth unilateral term with
the coat pattern of king cheetah and suggest a mechanism generating this pattern.

\subsection{{\bf Unilateral term yields irregular patterns}}
\begin{figure} 
\includegraphics[height=0.29\textwidth]{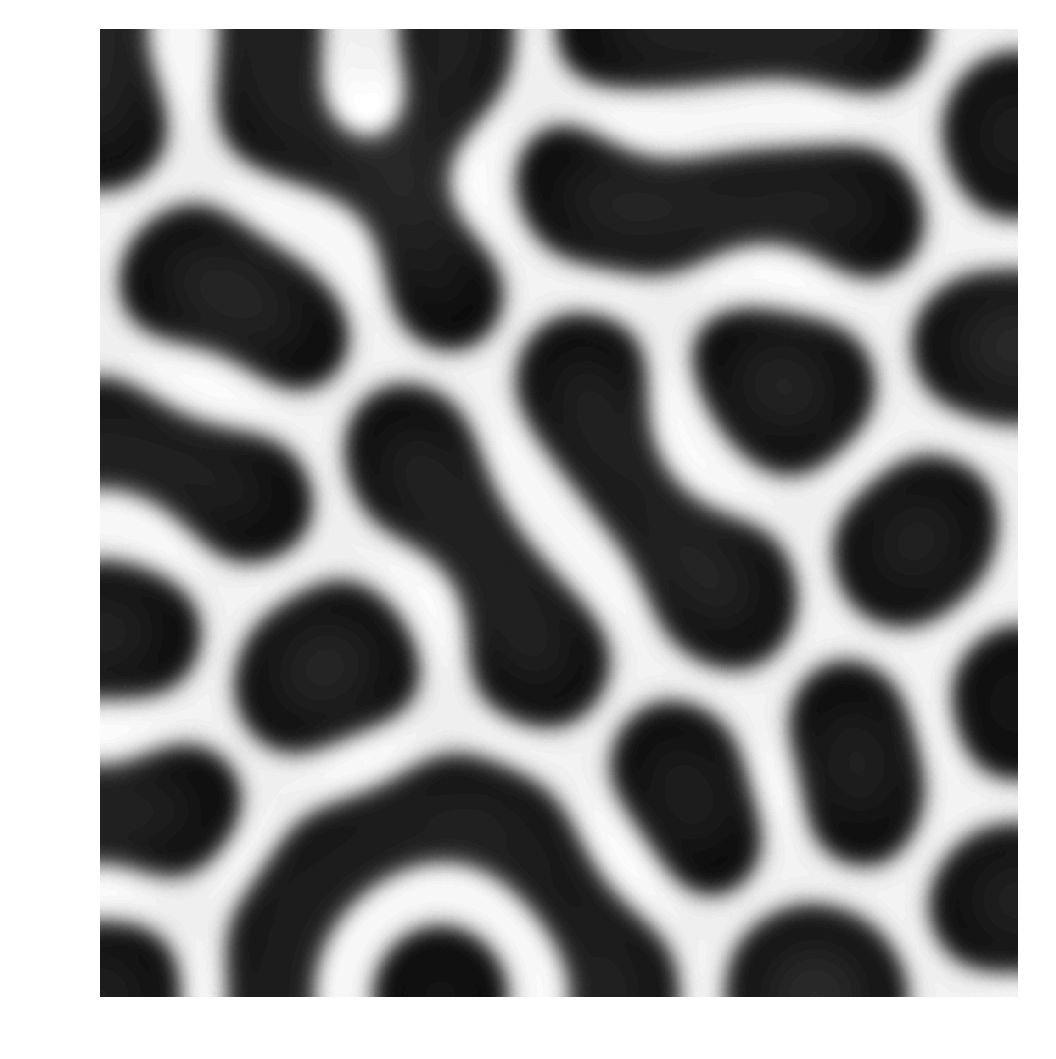} 
\includegraphics[height=0.29\textwidth]{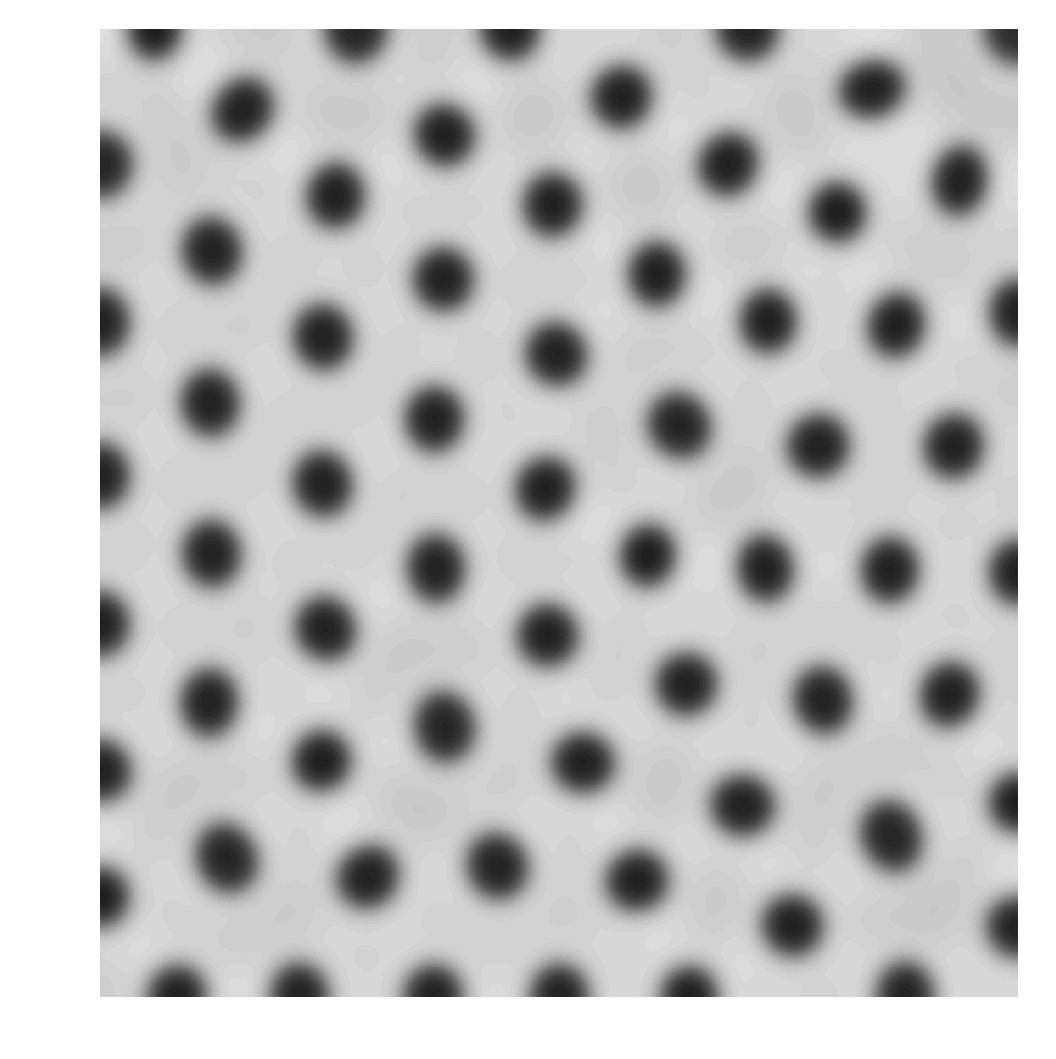} 
\includegraphics[height=0.29\textwidth]{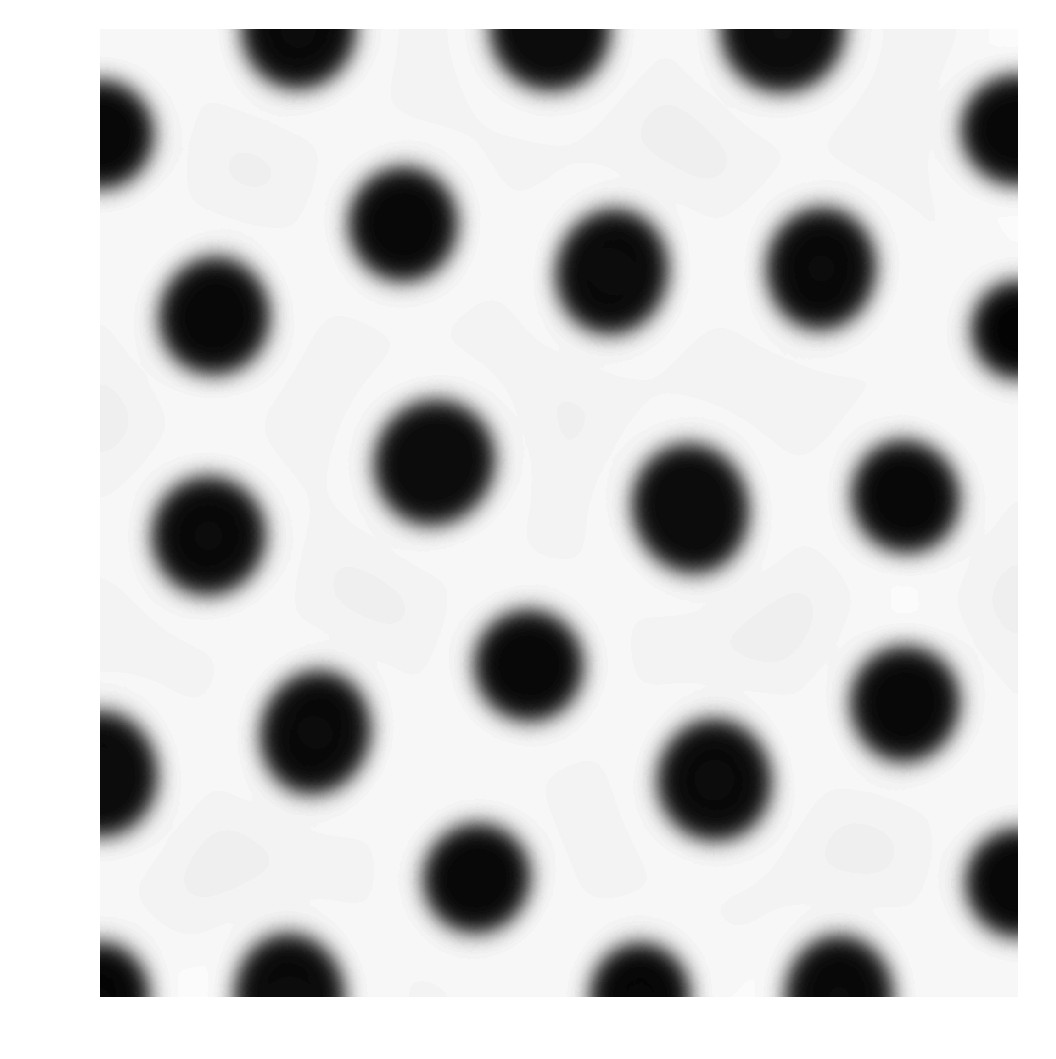} 
\includegraphics[height=0.29\textwidth]{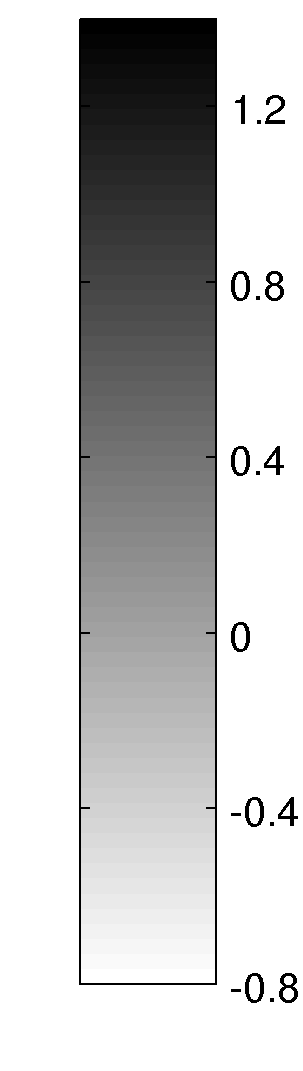} 
\caption{\label{fi:Maini1}
Typical patterns obtained by system \ttg{eq:Maini+g} with the nonsmooth unilateral term $\hat g(v) = \tau v^-$ (left panel), the classical case $\hat g \equiv 0$ (middle panel), and $\hat g(v) = -\tau v$ (right panel) with $\tau = 0.08$, $D=0.45$, and parameter values \ttg{eq:param}. The initial condition was specified as a small random noise around the ground state.
The grey scale shows the values of $u$.
}
\end{figure}

First, we compare the patterns produced by the nonsmooth unilateral term and by the linear terms $\hat g \equiv 0$ and $\hat g(v) = -\tau v$.
To this end we consider system \ttg{eq:Maini+g} with boundary conditions \ttg{eq:bc1}, and parameter values \ttg{eq:param}. Figure~\ref{fi:Maini1} compares patterns for choices 
$\hat g(v) = \tau v^-$, $\hat g \equiv 0$, and $\hat g(v) = -\tau v$, respectively,
for $\tau = 0.08$ and $D = 0.45$. 
%
Comparing these patterns 
we immediately observe the qualitative difference.
The linear choices of $\hat g(v)$
produce approximately circular spots which are, to some extent, symmetrically placed.
In contrast, the pattern produced by the unilateral system
shows irregular spots of larger size. Several of the largest spots seem
to be created by fusions of smaller spots. Moreover, the pattern
does not exhibit any symmetry even approximately.
Interestingly, similar irregular patterns are obtained in \cite{Woolley2010} by varying the parameter $h$ in the dimensionless version of model (3) without any unilateral source.

%

\subsection{{\bf Critical ratio of diffusions}}
Another interesting phenomenon resulting from the addition of the nonlinear unilateral source terms to the classical system (i.e. \ttg{eq:Maini+g} with $\hat g \equiv 0$) is 
the growth of small nonhomogeneous perturbations of the ground
state to patterns even if the ratio of diffusions exceeds the critical value
\ttg{eq:dcrit} of the classical system (i.e. $\hat g \equiv 0$). 
Indeed, the critical ratio 
of diffusions \ttg{eq:dcrit} for the classical system with parameter values \ttg{eq:param}
is $\dcrit \approx 0.53$.
However, using the nonsmooth unilateral source $\hat g(v) = \tau v^-$, 
we numerically obtain patterns forming
from very small spatial perturbations of the ground state even for considerably higher ratios
of diffusions. 
This phenomenon was predicted by a series of theoretical results, mainly \cite{SookVat,KucVat2012}.

\begin{figure} 
\includegraphics[width=0.9\textwidth]{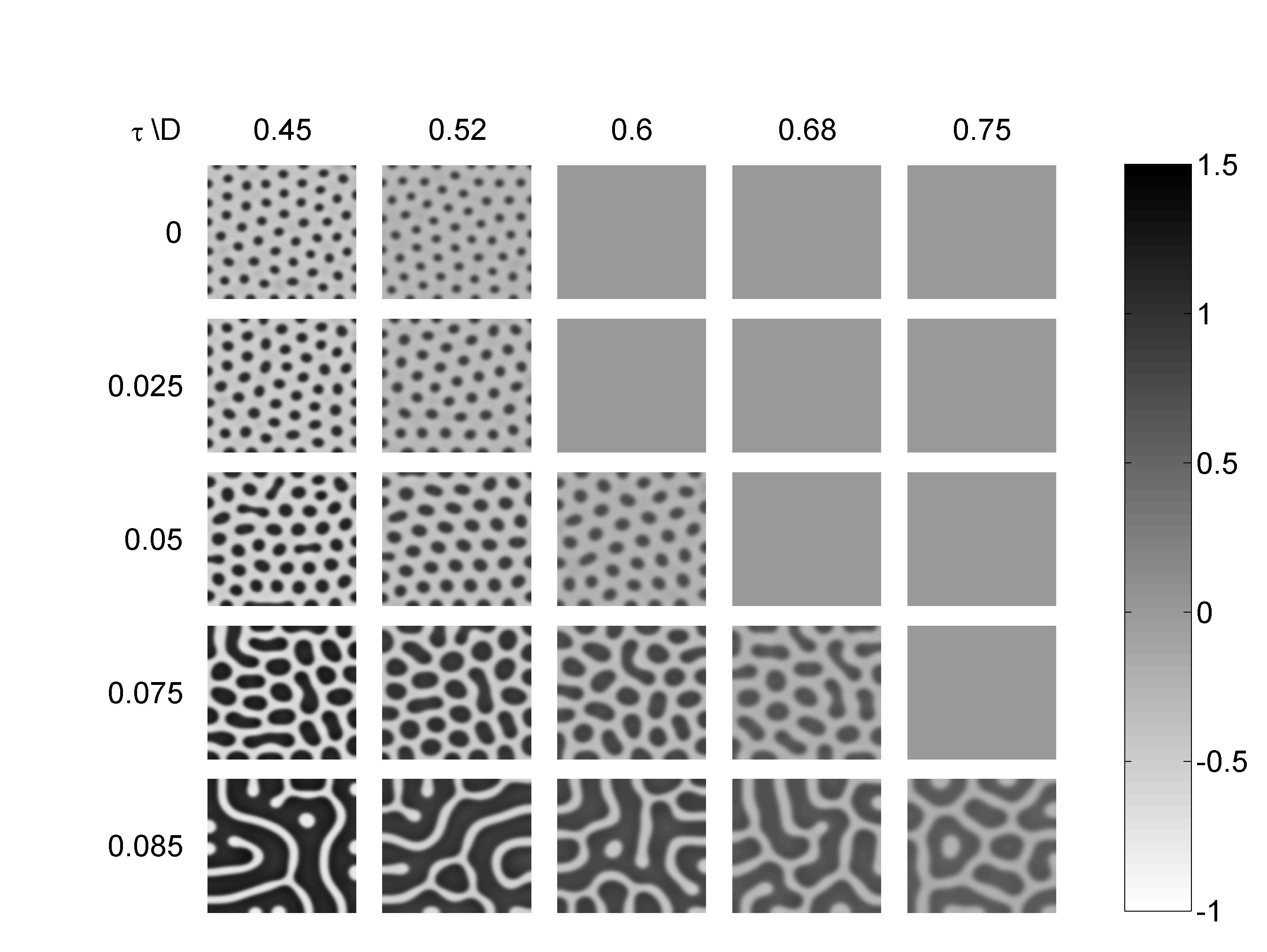} 



\caption{\label{fi:vzornik}
Dependence of patterns on the ratio of diffusions $D$ and the strength
of the unilateral source $\tau$ for the nonsmooth unilateral term $\hat g(v) = \tau v^-$.
Each box corresponds to the indicated values
of $D$ and $\tau$ and to parameter values \ttg{eq:param}.
}
\end{figure}

In order to illustrate the dependence of the arising patterns on
the strength of the unilateral source $\tau$ and on the ratio of diffusion
constants $D$, we present Figure~\ref{fi:vzornik}. The top-left box in
Figure~\ref{fi:vzornik} corresponds to the classical system ($\hat g \equiv 0$) with standard parameter values
\ttg{eq:param} and $D=0.45$. We observe the typical regular spotted pattern.
As $\tau$ increases, the spots are growing bigger and starting from certain value 
they seem to merge and irregular patterns emerge.
Similarly, we can observe that higher values of $\tau$ enable to produce
patterns for higher ratios of diffusions $D$.
In particular, columns 3--5 show that if $D$ exceeds the critical ratio of diffusions $\dcrit \approx 0.53$ of the classical system, then the spatial patterns arise only if
$\tau$ is sufficiently large. The larger is $D$, the larger $\tau$ is necessary for patterns to arise.
For completeness, we mention that numerically no patterns emerge for $\tau \geq 0.089$. 

\subsection{{\bf Linear term $\hat g=-\tau v$}}
\begin{figure} 
\includegraphics[width=0.9\textwidth]{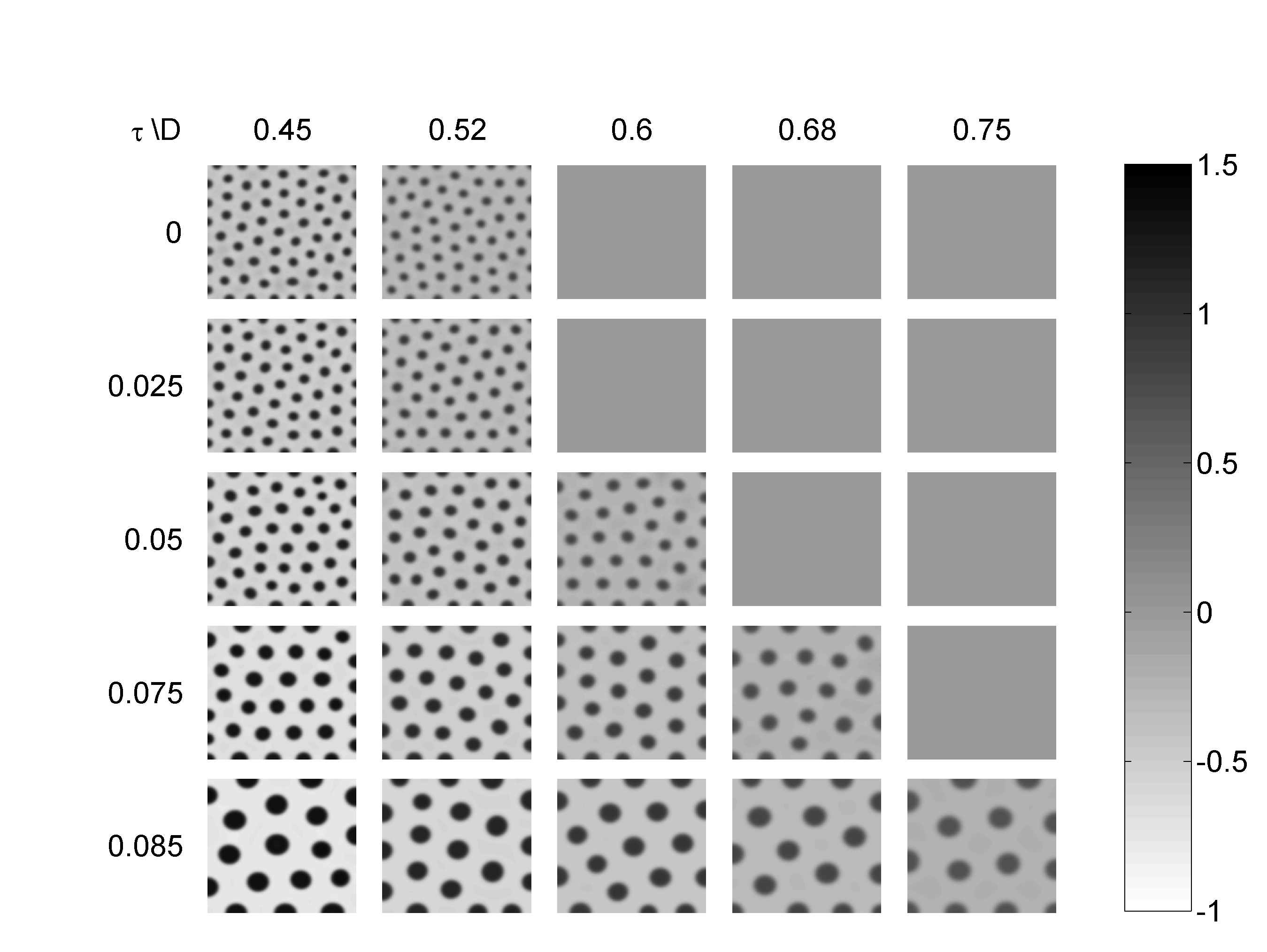}
\caption{\label{fi:vzornik2}
Dependence of patterns on $D$ and $\tau$ for the choice $\hat g(v) = -\tau v$.
Each box corresponds to the indicated values of $D$ and $\tau$ and to parameter values \ttg{eq:param}. 
}
\end{figure}

It is interesting to compare the above results with the case $\hat g(v) = -\tau v$.
Note that this choice can actually be seen as the classical system with $\hat g \equiv 0$ and coefficient $\beta$ modified to $\beta - \tau$. 
Figure~\ref{fi:vzornik2} shows the resulting patterns 
for various values of $D$ and $\tau$. This system is smooth and therefore we can analyse
the Turing instability including the critical ratios of diffusion coefficients
\ttg{eq:dcrit}.
Table~\ref{ta:Dcrit} presents these values for parameters \ttg{eq:param} and
various $\tau$.
Figure~\ref{fi:vzornik2} confirms that this system
produces patterns only if the ratio of diffusion coefficients is below
the critical value.
Interestingly, we observe patterns for the same values of the ratio of diffusions 
as for the system with $\hat g(v) = \tau v^-$ presented in
Figure~\ref{fi:vzornik}.
This leads us to a hypothesis that the Turing instability in the unilateral
system \ttg{eq:Maini+g} with $\hat g(v) = \tau v^-$ occurs under the same conditions as in the case of system \ttg{eq:Maini+g} with $\hat g(v) = -\tau v$.
%
Although, we do not present the results, we solved system \ttg{eq:Maini+g} with 
the nonsmooth unilateral term many times for values $D$ close to the critical one and all these
results confirmed this hypothesis.

On the other hand, comparison of Figures~\ref{fi:vzornik} and \ref{fi:vzornik2} clearly reveals the difference of the resulting patterns. The difference is even qualitative. While the patterns produced by the unilateral term are irregular with large irregular spots, patterns produced by the linear term are approximately symmetric with smaller circular spots. This qualitative difference can be explained by the substantial difference of the corresponding nonlinear dynamics especially for values of $v$ distant from the ground state.

Another difference can be visible if we observe how the patterns change as $\tau$ is increased. In both cases the spots grow bigger, but differently. For the unilateral term the spots grow, fill gaps among them, and finally merge. On the other hand, for the linear term the distances of spots grow proportionally and spots do not merge. This behaviour is similar as if we scaled the system by the size of the domain $\Omega$ or by parameter $\delta$. We note that in the dimensionless system \cite{Aragon2012} (where no unilateral term is considered) a parameter $\eta$ scales the system in the same manner.

\begin{table}[b]
\begin{tabular}{c|cccccc}
$\tau$ & 0 & 0.025 & 0.05 & 0.075 & 0.08 & 0.085 
\\ \hline
$\dcrit$ & 0.53 & 0.57 & 0.62 & 0.71 & 0.74 & 0.78
\end{tabular}
\caption{\label{ta:Dcrit}
Critical ratios \ttg{eq:dcrit} of diffusion coefficients for the linear source term (i.e. system \ttg{eq:Maini+g} with $\hat g(v) = -\tau v$)
and various values of $\tau$. Rounded to two significant digits.
}
\end{table}

\subsection{{\bf  Unilateral source as a model of a receptor-based morphogen regulation}}
\label{se:modelmorph}
It is usually considered in reaction-diffusion models of pre-pattern formation in mammalian skin that morphogens are proteins (ligands) secreted to the extracellular 
space \cite{HeaPaiSip2009}. These proteins do not react directly with each other but they 
bind to cell membrane receptors and the production of morphogens is 
subsequently regulated by signaling pathways. The mechanism of cell response to morphogen gradients is a subject of intensive debates \cite{SagBri2017, AshBri2006}. It is assumed here that the number of receptors engaged with ligands influences the rate of morphogen production \cite{GouBou2001} and that this influence is in inverse proportion. The introduction of a unilateral term to our model reflects a limited number of receptors in a membrane: if the concentration of a morphogen exceeds the threshold value $\theta$, all receptors are occupied independently on the amount how much the threshold is exceeded. The 
system is saturated and does not produce the corresponding morphogen.
In the case of the morphogen $v$, this process is well described by
the term $\hat g(v) = \tau (v-\theta)^-$.
Indeed, in points $(x,y)\in\Omega$ and times $t$, where $v(x,y,t) < \theta$,
the term $\tau (v(x,y,t)-\theta)^-$
is positive and works as a source term in \ttg{eq:zdroj}. On the other hand,
in points $(x,y)\in\Omega$ and times $t$, where $v(x,y,t) \geq \theta$,
the term $\tau (v(x,y,t)-\theta)^-$ vanishes and has no effect.

\subsection{{\bf Case study: Coat patterns of cheetahs}}
\label{se:kingcheetah}

%

\begin{figure} 
\includegraphics[width=0.45\textwidth]{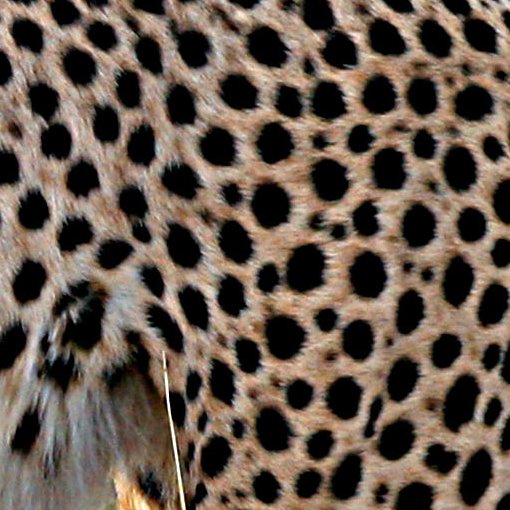}\quad
\includegraphics[width=0.45\textwidth]{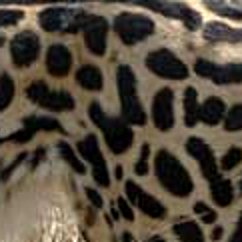}
\caption{\label{fi:cheetah}
Typical coat patterns of the common morph of the cheetah (left) and king cheetah (right).
}
\end{figure}

It has been shown that \emph{Taqpep} gene is responsible for the regularity of pre-pattern in the
case of domestic cats and cheetahs \cite{Kaelin2012}, see Figure~\ref{fi:cheetah} (left). King cheetahs have a mutation in this gene and their specific coat pattern is characterized by irregular, large spots, see Figure~\ref{fi:cheetah} (right). \emph{Taqpep} encodes a type II membrane-spanning protein of the M1 aminopeptidase family whose metalloprotease-containing ectodomain (further denoted as MCE) can diffuse outside the cell.

It has been proposed that a reaction-diffusion model is suitable to elucidate a role of MCE for the constitution of pre-patterns \cite{Kaelin2013}. 
Mathematical model \eqref{eq:Maini+g} and considerations from Section~\ref{se:modelmorph} serves this purpose, presuming that variable $v$ is the deviation of the MCE concentration in the extracellular space from its equilibrium concentration $\bar V$. 

According to the model, the unilateral regulation is weak in the common morph of cheetah ($\tau$ close to zero), resulting in the usual spotted pattern, whereas the mutation in \emph{Taqpep} gene yields stronger unilateral regulation in the case of the king cheetah ($\tau$ around 0.08). The simulated patterns relate to real skin patterns, which applies both for the common morph of cheetah, see Figures~\ref{fi:Maini1} (middle) and \ref{fi:cheetah} (left), and the king cheetah, see Figures~\ref{fi:Maini1} (left) and \ref{fi:cheetah} (right).

\section{Discussion and conclusions}
\label{se:conslusion}

In this contribution we investigated a reaction-diffusion system with
a nonsmooth unilateral source term $\hat g(v) = \tau v^-$
and its approximations.
We provided a case study for a particular system and 
analysed numerically the influence of such source term on 
the Turing instability and on the resulting patterns.
We explained a possible biological meaning of this term and obtained the following conslusions.

{\bf Sensitivity of the Turing instability.}
The linear analysis  
of systems with smooth approximations of the term $\hat g(v) = \tau v^-$ 
is not informative about the Turing instability of the system with the nonsmooth term. 
Small perturbations of the ground state can evolve to patterns for one 
approximation of the nonsmooth term, but not for the other even though they are arbitrarily accurate. This shows that the Turing instability is sensitive to small changes of the nonlinear dynamics.

{\bf Robustness of Turing patterns.}
We also showed that initial perturbations of the ground state 
larger than a certain minimal size do robustly evolve to patterns for both the nonsmooth term
and its approximations. In addition,
these patterns are almost identical regardless the particular 
form of the unilateral term
in the small neighbourhood of the ground state.

{\bf Irregularity of patterns.}
We have found that
unilateral sources break the approximate regularity and symmetry of
the usual patterns.
System \ttg{eq:Maini+g} with the unilateral term
$\hat g(v) = \tau v^-$ produces
spots with irregular shapes and variable distances between them.
This contrasts to the classical smooth systems corresponding to choices
$\hat g \equiv 0$ and $\hat g(v) = -\tau v$ in \ttg{eq:Maini+g}, where
we observe close-to-regular disc-shaped spots approximately symmetrically placed,
see Figure~\ref{fi:Maini1}.
Thus, the unilateral sources
prescribed for the inhibitor $v$ break the regularity
of patterns for all values of diffusion constants yielding patterns,
provided the strength of the unilateral source is not negligible.

{\bf Patterns for higher ratio of diffusion parameters.}
Interestingly, system \ttg{eq:Maini+g} with the unilateral term $\hat g(v) = \tau v^-$
produces patterns even
for those values of diffusion constants which prevent any pattern formation in
the original system (i.e. $\hat g \equiv 0$).
Further, we observe that the critical ratio of diffusions for the
system with $\hat g(v) = \tau v^-$ seems to be identical to the critical ratio of the system with $\hat g(v) = -\tau v$. However, the resulting patterns differ considerably as we mentioned in the previous paragraph.

\smallskip

We verified these conclusions numerically for the particular system \ttg{eq:Maini+g}, but we believe that they are valid for other kinetics as well. 
We verified this conclusion in \cite{RybVej2014} for a unilateral term added to the Thomas model \cite{thomas}. This indicates that our findings about the effects of the unilateral term are not limited to a single kinetics. The generality of these results is also supported by earlier results \cite{DKM,SookVat,KucVat2012}, which guarantee the existence of bifurcations of spatial patterns and certain instability of the ground state for large $d_1/d_2$ for a unilateral source described by variational inequalities, but they are valid for a very large class of kinetics.

Reaction-diffusion systems with nonsmooth nonlinear unilateral terms are interesting from both
the theoretical and practical points of view.
In contrasts to the classical smooth
case, where the small perturbations initially evolve according to a linear dynamics,
the evolution of small perturbations of the ground state for the nonsmooth unilateral term
is inherently governed by a nonlinear dynamics. 
This nonlinear dynamics
may yield completely new phenomena in the pattern formation mechanisms.
In this contribution, we have made an attempt towards the understanding
of the unilateral terms 
in models of biological patterns formation. 
However, further research is necessary for the investigation of feasible biological applications. 
%


\section*{ACKNOWLEDGMENTS}
The research leading to these results has received funding from the People Programme (Marie Curie Actions) of the European Union's Seventh Framework Programme (FP7/2007-2013) under REA grant agreement no. 328008.
Further, M.K. has been supported by the Grant 13-00863S of the Czech Science Foundation and
T.V., M.K., and V.R. acknowledge the support of RVO~67985840.


\begin{thebibliography}{10}
\providecommand{\url}[1]{{#1}}
\providecommand{\urlprefix}{URL }
\expandafter\ifx\csname urlstyle\endcsname\relax
  \providecommand{\doi}[1]{DOI~\discretionary{}{}{}#1}\else
  \providecommand{\doi}{DOI~\discretionary{}{}{}\begingroup
  \urlstyle{rm}\Url}\fi

\bibitem{Aragon2012}
Arag{\'o}n, J., Barrio, R., Woolley, T., Baker, R., Maini, P.: Nonlinear
  effects on {T}uring patterns: {T}ime oscillations and chaos.
\newblock Physical Review E \textbf{86}(2), 026,201 (2012)

\bibitem{AshBri2006}
Ashe, H.L., Briscoe, J.: The interpretation of morphogen gradients.
\newblock Development \textbf{133}(3), 385--394 (2006)

\bibitem{Asslani2012}
Asslani, M., Di~Patti, F., Fanelli, D.: Stochastic turing patterns on a
  network.
\newblock Physical Review E \textbf{86}(4), 046,105 (2012)

\bibitem{BarVarAra1999}
Barrio, R., Varea, C., Arag{\'o}n, J., Maini, P.: A two-dimensional numerical
  study of spatial pattern formation in interacting turing systems.
\newblock Bulletin of mathematical biology \textbf{61}(3), 483--505 (1999)

\bibitem{Biancalani2010}
Biancalani, T., Fanelli, D., Di~Patti, F.: Stochastic turing patterns in the
  brusselator model.
\newblock Physical Review E \textbf{81}(4), 046,215 (2010)

\bibitem{Biancalani2017}
Biancalani, T., Jafarpour, F., Goldenfeld, N.: Giant amplification of noise in
  fluctuation-induced pattern formation.
\newblock Physical review letters \textbf{118}(1), 018,101 (2017)

\bibitem{Cantini2014}
Cantini, L., Cianci, C., Fanelli, D., Massi, E., Barletti, L., Asllani, M.:
  Stochastic amplification of spatial modes in a system with one diffusing
  species.
\newblock Journal of mathematical biology \textbf{69}(6-7), 1585--1608 (2014)

\bibitem{Chen2001}
Chen, Y., Schier, A.F.: The zebrafish nodal signal squint functions as a
  morphogen.
\newblock Nature \textbf{411}(6837), 607--610 (2001)

\bibitem{Ciarlet1978}
Ciarlet, P.G.: The finite element method for elliptic problems.
\newblock North-Holland Publishing Co., Amsterdam-New York-Oxford (1978).
\newblock Studies in Mathematics and its Applications, Vol. 4

\bibitem{DKM}
Dr{\'a}bek, P., Ku{\v{c}}era, M., M{\'\i}kov{\'a}, M.: Bifurcation points of
  reaction-diffusion systems with unilateral conditions.
\newblock Czechoslovak Mathematical Journal \textbf{35}(4), 639--660 (1985)

\bibitem{Economou2012}
Economou, A.D., Ohazama, A., Porntaveetus, T., Sharpe, P.T., Kondo, S., Basson,
  M.A., Gritli-Linde, A., Cobourne, M.T., Green, J.B.: Periodic stripe
  formation by a {T}uring mechanism operating at growth zones in the mammalian
  palate.
\newblock Nature genetics \textbf{44}(3), 348--351 (2012)

\bibitem{Edelstein1988}
Edelstein-Keshet, L.: Mathematical models in biology.
\newblock McGraw-Hill, Boston (1988)

\bibitem{EisKuc}
Eisner, J., Ku{\v{c}}era, M.: Spatial patterning in reaction-diffusion systems
  with nonstandard boundary conditions.
\newblock Fields Inst. Commun. \textbf{25}, 239--256 (2000)

\bibitem{EisKuc1999}
Eisner, J., Ku{\v{c}}era, M.: Bifurcation of solutions to reaction-diffusion
  systems with jumping nonlinearities.
\newblock In: A.~Sequeira, H.~Beirao~da Veiga, J.H. Videman (eds.) Applied
  Nonlinear Analysis, pp. 79--96. Springer (2002)

\bibitem{EisVat2011}
Eisner, J., V{\"a}th, M.: Location of bifurcation points for a
  reaction-diffusion system with {N}eumann-{S}ignorini conditions.
\newblock Advanced Nonlinear Studies \textbf{11}(4), 809--836 (2011)

\bibitem{Fanelli2013}
Fanelli, D., Cianci, C., Di~Patti, F.: Turing instabilities in
  reaction-diffusion systems with cross diffusion.
\newblock The European Physical Journal B \textbf{86}(4), 142 (2013)

\bibitem{Filippov:1988}
Filippov, A.: Differential Equations with Discontinuous Righthand Sides.
\newblock Kluwer Academic Publishers (1988)

\bibitem{GouBou2001}
Gurdon, J., Bourillot, P.Y.: Morphogen gradient interpretation.
\newblock Nature \textbf{413}(6858), 797--803 (2001)

\bibitem{HeaPaiSip2009}
Headon, D.J., Painter, K.J.: Stippling the skin: Generation of anatomical
  periodicity by reaction-diffusion mechanisms.
\newblock Mathematical Modelling of Natural Phenomena \textbf{4}(4), 83--102
  (2009)

\bibitem{JonesSleeman}
Jones, D.S., Sleeman, B.D.: Differential Equations and Mathematical Biology.
\newblock Chapman \& Hall (2003)

\bibitem{Kaelin2013}
Kaelin, C.B., Barsh, G.S.: Genetics of pigmentation in dogs and cats.
\newblock Annu. Rev. Anim. Biosci. \textbf{1}(1), 125--156 (2013)

\bibitem{Kaelin2012}
Kaelin, C.B., Xu, X., Hong, L.Z., David, V.A., McGowan, K.A.,
  Schmidt-K{\"u}ntzel, A., Roelke, M.E., Pino, J., Pontius, J., Cooper, G.M.,
  et~al.: Specifying and sustaining pigmentation patterns in domestic and wild
  cats.
\newblock Science \textbf{337}(6101), 1536--1541 (2012)

\bibitem{SookVat}
Kim, I.S., V{\"a}th, M.: The {K}rasnoselskii-{Q}uittner formula and instability
  of a reaction-diffusion system with unilateral obstacles.
\newblock Dynamics of Partial Differential Equations \textbf{11}(3), 229--250
  (2014)

\bibitem{Klika2012}
Klika, V., Baker, R.E., Headon, D., Gaffney, E.A.: The influence of
  receptor-mediated interactions on reaction-diffusion mechanisms of cellular
  self-organisation.
\newblock Bulletin of mathematical biology \textbf{74}(4), 935--957 (2012)

\bibitem{Kondo2002}
Kondo, S.: The reaction-diffusion system: a mechanism for autonomous pattern
  formation in the animal skin.
\newblock Genes to Cells \textbf{7}(6), 535--541 (2002)

\bibitem{Kondo2010}
Kondo, S., Miura, T.: Reaction-diffusion model as a framework for understanding
  biological pattern formation.
\newblock Science \textbf{329}(5999), 1616--1620 (2010)

\bibitem{KucVat2012}
Ku{\v{c}}era, M., V{\"a}th, M.: Bifurcation for a reaction--diffusion system
  with unilateral and {N}eumann boundary conditions.
\newblock Journal of Differential Equations \textbf{252}(4), 2951--2982 (2012)

\bibitem{Kus2015}
K{\r{u}}s, P.: Convergence and stability of higher-order finite element
  solution of reaction-diffusion equation with turing instability.
\newblock In: J.~Brandts, S.~Korotov, M.~K\v{r}\'\i\v{z}ek, K.~Segeth,
  J.~\v{S}{\'\i}stek, T.~Vejchodsk\'y (eds.) Application of Mathematics 2015,
  pp. 140--147. Institute of Mathematics CAS (2015)

\bibitem{KucBos1994}
Ku\v{c}era, M., Bos{\'a}k, M.: Bifurcation for quasi-variational inequalities
  of reaction-diffusion type.
\newblock Stability and Applied Analysis of Continuous Media \textbf{3}(2),
  354--369 (1994)

\bibitem{LiuLiaMai2006}
Liu, R., Liaw, S., Maini, P.: Two-stage {T}uring model for generating pigment
  patterns on the leopard and the jaguar.
\newblock Physical review E \textbf{74}(1), 01{19}14 (2006)

\bibitem{Mou2006}
Mou, C., Jackson, B., Schneider, P., Overbeek, P.A., Headon, D.J.: Generation
  of the primary hair follicle pattern.
\newblock Proceedings of the National Academy of Sciences \textbf{103}(24),
  9075--9080 (2006)

\bibitem{Murray2003}
Murray, J.D.: Mathematical biology. {II}. {S}patial models and biomedical
  applications.
\newblock Springer-Verlag (2003)

\bibitem{Nishiura1982}
Nishiura, Y.: Global structure of bifurcating solutions of some
  reaction-diffusion systems.
\newblock SIAM Journal on Mathematical Analysis \textbf{13}(4), 555--593 (1982)

\bibitem{RybVej2014}
Ryb\'a\v{r}, V., Vejchodsk\'y, T.: Irregularity of {T}uring patterns in the
  {T}homas model with a unilateral term.
\newblock In: J.~Chleboun, P.~P\v{r}ikryl, K.~Segeth, J.~\v{S}\'{\i}stek,
  T.~Vejchodsk\'y (eds.) Programs and Algorithms of Numerical Matematics 17,
  pp. 188--193. Institute of Mathematics AS CR (2015)

\bibitem{SagBri2017}
Sagner, A., Briscoe, J.: Morphogen interpretation: concentration, time,
  competence, and signaling dynamics.
\newblock Wiley Interdisciplinary Reviews: Developmental Biology \textbf{6},
  e271 (2017).
\newblock \doi{10.1002/wdev.271}

\bibitem{Schier2009}
Schier, A.F.: Nodal morphogens.
\newblock Cold Spring Harbor perspectives in biology \textbf{1}(5), a00{34}59
  (2009)

\bibitem{thomas}
Thomas, D.: Artificial enzyme membranes, transport, memory, and oscillatory
  phenomena.
\newblock In: D.~Thomas, J.P. Kernevez (eds.) Analysis and control of
  immobilized enzyme systems, pp. 115--150. North Holland (1976)

\bibitem{Turing1952}
Turing, A.M.: The chemical basis of morphogenesis.
\newblock Philosophical Transactions of the Royal Society of London B:
  Biological Sciences \textbf{237}(641), 37--72 (1952)

\bibitem{Woolley2010}
Woolley, T.E., Baker, R.E., Maini, P.K., Arag{\'o}n, J.L., Barrio, R.A.:
  Analysis of stationary droplets in a generic turing reaction-diffusion
  system.
\newblock Physical Review E \textbf{82}(5), 051,929 (2010)

\end{thebibliography}
\end{document}